\documentclass[fleqn,usenatbib]{mnras}

\usepackage{newtxtext,newtxmath}

\usepackage[T1]{fontenc}

\DeclareRobustCommand{\VAN}[3]{#2}
\let\VANthebibliography\thebibliography
\def\thebibliography{\DeclareRobustCommand{\VAN}[3]{##3}\VANthebibliography}

\usepackage{graphicx}	
\usepackage{amsmath}	

\newcommand{\fsn}{f_\text{SN}}
\newcommand{\vsn}{\Delta v_\text{SN}}
\newcommand{\tagn}{\Delta T_\text{AGN}}
\newcommand{\betabh}{\beta_\text{BH}}

\newcommand{\kms}{\text{km}\,\text{s}^{-1}}

\title[Hydrostatic mass bias for FLAMINGO clusters]{Hydrostatic mass bias for galaxy groups and clusters in the FLAMINGO simulations}

\author[J. Braspenning et al.]{
Joey Braspenning,$^{1}$\thanks{E-mail: braspenning@strw.leidenuniv.nl}
Joop Schaye,$^{1}$
Matthieu Schaller,$^{2, 1}$
Roi Kugel,$^{1}$
Scott T. Kay$^{3}$
\\
$^{1}$Leiden Observatory, Leiden University, PO Box 9513, 2300 RA Leiden, the Netherlands\\
$^{2}$Lorentz Institute for Theoretical Physics, Leiden University, PO box 9506, 2300 RA Leiden, the Netherlands\\
$^{4}$Jodrell Bank Centre for Astrophysics, Department of Physics and Astronomy, The University of Manchester, Oxford Road, Manchester M13 9PL, UK\\
}

\date{Accepted XXX. Received YYY; in original form ZZZ}

\pubyear{2024}

\begin{document}
\label{firstpage}
\pagerange{\pageref{firstpage}--\pageref{lastpage}}
\maketitle

\begin{abstract}
The masses of galaxy clusters are commonly measured from X-ray observations under the assumption of hydrostatic equilibrium (HSE). This technique is known to underestimate the true mass systematically. The fiducial FLAMINGO cosmological hydrodynamical simulation predicts the median hydrostatic mass bias to increase from $b_\text{HSE} \equiv (M_\text{HSE,500c}-M_\text{500c})/M_\text{500c} \approx -0.1$ to -0.2 when the true mass increases from group to cluster mass scales. However, the bias is nearly independent of the hydrostatic mass. The scatter at fixed true mass is minimum for $M_\text{500c}\sim 10^{14}~\text{M}_\odot$, where $\sigma(b_\text{HSE})\approx 0.1$, but increases rapidly towards lower and higher masses. At a fixed true mass, the hydrostatic masses increase (decrease) with redshift on group (cluster) scales, and the scatter increases. The bias is insensitive to the choice of analytic functions assumed to represent the density and temperature profiles, but it is sensitive to the goodness of fit, with poorer fits corresponding to a stronger median bias and a larger scatter. The bias is also sensitive to the strength of stellar and AGN feedback. Models predicting lower gas fractions yield more (less) biased masses for groups (clusters). The scatter in the bias at fixed true mass is due to differences in the pressure gradients rather than in the temperature at $R_\text{500c}$. The total kinetic energies within $r_\text{500c}$ in low- and high-mass clusters are sub- and super-virial, respectively, though all become sub-virial when external pressure is accounted for. Analyses of the terms in the virial and Euler equations suggest that non-thermal motions, including rotation, account for most of the hydrostatic mass bias. However, we find that the mass bias estimated from X-ray luminosity weighted profiles strongly overestimates the deviations from hydrostatic equilibrium.
\end{abstract}

\begin{keywords}
galaxies: clusters: intracluster medium -- large-scale structure of Universe -- galaxies: clusters: general -- methods: numerical -- X-rays: galaxies: clusters
\end{keywords}



\section{Introduction}

Measuring masses of massive groups and clusters of galaxies poses a formidable challenge because there is no universally applicable or unbiased method. Whereas simulations can simply measure the mass of all the particles or volume elements enclosed within a given 3D radius, observations do not have that luxury. Two widely used methods are estimating masses from X-ray observations under the assumption of hydrostatic equilibrium and weak gravitational lensing observations, both of which have their own challenges. 

Weak lensing observations can be used to infer masses by measuring the slight distortion of background objects around massive clusters. Masses measured in this way are inherently projected along the line of sight, polluting the mass estimate by correlated fore-and-background structures \citep[e.g.][]{Debackere2022}. 

Uncertainties associated with weak lensing masses consist mainly of three parts. Shape measurements can be inaccurate due to the difficulty of correcting the anisotropic PSF \citep[e.g.][]{Hamana2013}, which is most important for the shapes of faint galaxies \citep[e.g.][]{Heymans2006, Massey2007, Bridle2010, Kitching2012, Kitching2013, Mandelbaum2015}. Background galaxy selection can dilute cluster cores, which depends on redshift, richness, and radius \citep[e.g.][]{OkabeSmith2016}, correcting for this can change the mass estimates by up to 40 per cent but is highly uncertain \citep{Okabe2016}. Finally, the weak lensing mass inference depends strongly on the orientation of the halo \citep[e.g.][]{Herbonnet2022}. For example, \citet{Munoz2024} find that the true mass is under-estimated by 25 per cent when aligned with the second principal axis of the cluster, but over-estimated by the same amount when aligned with the zeroth principal axis. 

Alternatively, the mass can be estimated using the hydrostatic mass inferred from X-ray observations. Using the surface brightness distribution, the gas density profile can be extracted \citep[e.g.][]{Pratt2002, Ettori2013, Ghirardini2019}. Assuming an emission model, the spatially resolved X-ray spectrum can be converted into a temperature profile \citep[e.g.][]{Arnaud1996}. Under the assumption of hydrostatic equilibrium, fits to these profiles with an analytic function can be used to compute the enclosed mass \citep[see e.g.][]{Sarazin1988}.

In practice, hydrostatic mass measurements pose a significant challenge. To obtain the mass enclosed within a sphere, the density and temperature profiles have to be three-dimensional, which means observations have to be deprojected from the plane of the sky in which they are observed \citep[see e.g.][for a comprehensive description of the process]{Turner2024}. Unlike in the case for weak lensing mass estimates, the cluster shape only has a 1 per cent effect on the mass estimate \citep{BuoteHumphrey2012}. However, temperature inhomogeneities within a single radial bin of the profile can have a 10 per cent effect on the mass, due to clumpy cold gas biasing the mean temperature low \citep{Mazzotta2004, Rasia2014, Henson2017}. Forward modelling and external observations, e.g. of the Sunyaev-Zeldovich effect, can improve the hydrostatic mass measurements and the estimates of the associated errors.

On top of that, to measure the total mass of a galaxy cluster, observations have to extend quite far out from the centre to capture the full extent of the cluster's mass distribution. Even at $R_{\rm 500c}$\footnote{$R_{\rm 500c}$ is the radius at which the average interior density is 500 times the critical density of the Universe.} the X-ray surface brightness tends to be very low and observations carry large error bars. As the mass depends on the spatial derivative of the profiles at a given radius, small errors in the profiles can translate to large errors in the mass estimates. For example, \citet{Rossetti2024} find that a careful re-analysis of their galaxy cluster data changes the slope of the temperature profile by 20 per cent.

The total hydrostatic mass is linearly dependent on the absolute temperature. In observations, there is a known systematic difference of up to 30 per cent between the temperatures inferred from the eROSITA, XMM-\textit{Newton} and \textit{Chandra} telescopes, especially above $5~\mathrm{keV}$ \citep{Migkas2024, Schellenberger2015}. However, \citet{Martino2014} find near perfect agreement in the inferred mass between XMM-\textit{Newton} and \textit{Chandra}, suggesting that the slopes of the retrieved density and temperature profiles also differ to compensate for the offset in temperature.

The hydrostatic masses inferred from X-ray observations are thus found to be biased. One method of determining the hydrostatic mass bias is by comparing the X-ray inferred masses ($M_{\rm 500c, X}$) with weak lensing masses ($M_{\rm 500c, WL}$). Using this approach, the XXL cluster survey found a value of $M_{\rm X} / M_{\rm WL} = 0.72 \pm 0.08$ for $2\times10^{14}~\mathrm{M_\odot} < M_{\rm 500c} < 10^{15}~\mathrm{M_\odot}$ \citep{Eckert2016}, and the Canadian Cluster Project in combination with the Multi Epoch Nearby Cluster Survey obtained $M_{\rm X} / M_{\rm WL} = 0.84 \pm 0.04$ for $4\times10^{14}~\mathrm{M_\odot} < M_{\rm 500c} < 2\times10^{15}~\mathrm{M_\odot}$ \citep{Herbonnet2020}, a slightly smaller bias than $M_{\rm X} / M_{\rm WL} = 0.74 \pm 0.06$ measured by \citet{Lovisari2020_mass} who expanded the measurements with new X-ray and weak lensing observations in the mass range $3\times10^{14}~\mathrm{M_\odot} < M_{\rm 500c} < 2\times10^{15}~\mathrm{M_\odot}$. However, much weaker biases of $M_{\rm X} / M_{\rm WL} = 0.95 \pm 0.05$ for $2\times10^{14}~\mathrm{M_\odot} < M_{\rm 500c} < 10^{15}~\mathrm{M_\odot}$ were found by \citet{Smith2016} and later $M_{\rm X} / M_{\rm WL} = 0.84 \pm 0.095$ for $3\times10^{14}~\mathrm{M_\odot} < M_{\rm 500c} < 10^{15}~\mathrm{M_\odot}$  by \citet{Wicker2023}.

In simulations, the hydrostatic mass can be compared to the true mass. \citet{Barnes2021} applied a consistent forward modeling pipeline and hydrostatic mass inference technique to a diverse set of cosmological simulations, yielding $M_{\rm 500c}$ bias measurements for $M_{\rm 500c} \geq 10^{14}~\mathrm{M_\odot}$ of $M_{\rm X} / M_{\rm True} = 0.89 \pm 0.01$ for IllustrisTNG \citep{Springel2018}, $M_{\rm X} / M_{\rm True} = 0.87 \pm 0.002$ for BAHAMAS \citep{McCarthy2017}, and $M_{\rm X} / M_{\rm True} = 0.85 \pm 0.003$ for the MACSIS \citep{Barnes2017_macsis} simulations. \citet{Ansarifard2020} found that in the 300 simulations (which yield $M_{\rm X} / M_{\rm True} = 0.9 \pm 0.09$), the bias can be accounted for by correcting for the azimuthal scatter and thermal pressure slope derived from the mock X-ray emission. \citet{Jennings2023} find a bias that is $M_{\rm X} / M_{\rm True} = 0.85 \pm 0.15$ for mass-weighted profiles, but $M_{\rm X} / M_{\rm True} = 0.7 \pm 0.15$ for emission-weighted profiles for the SIMBA simulations. However, compared to weak lensing - X-ray cluster samples, the simulated clusters are, on average, of lower mass. While all cosmological simulations that were compared by \citet{Barnes2021} showed a bias increasing with mass, the limited number of simulated clusters with $M_{\rm 500c} \approx 10^{15}~\mathrm{M_\odot}$ makes reliable comparisons challenging.

Among the explanations for the hydrostatic bias is the neglected contribution from non-thermal pressure \citep[e.g.][]{Kay2004, Nagai2007, Martizzi2016, Ettori2022}. One important source of non-thermal pressure could be contributions from magnetic fields and cosmic rays \citep[see][for a review]{Rusdkowski2023}. However, \citet{Biffi2016} argued that the object-to-object variation is far too large to effectively include a correction term in the hydrostatic mass calculation. Furthermore, in a joint analysis of Sunyaev–Zel'dovich (SZ) effect observations, strong-lensing data, and weak-lensing data, \citet{Siegel2018} constrained the non-thermal pressure fraction to be $< 0.11$, indicating that it would at most be a modest correction. However, that number is significantly lower than the turbulent pressure contribution found in some simulations, which can exceed 30 per cent \citep{Nelson2014}. Future observations with high-resolution X-ray spectrographs will resolve the spatial turbulence, potentially leading to self-consistent corrections to the hydrostatic mass on an object-by-object basis \citep[e.g.][]{Ota2018}.
Apart from the contribution of non-thermal pressure, the hydrostatic mass bias can also arise if clusters are out of equilibrium, e.g.\ due to recent mergers.

Fitting analytic profiles, based on the expected shape of the cluster density distribution, can help overcome low signal-to-noise at large radii. Because the analytic profiles are fit to the entire range of observed densities and temperatures, the higher fidelity data nearer to the cluster centre will contribute to the constraints, resulting in a more precise mass estimate. The results will, however, only be more accurate if the chosen profile is realistic.

Recent simulations by \citet{Scheck2023} have furthermore shown that even in idealised hydrostatic clusters, there is a bias of up to 7 per cent in the retrieved hydrostatic mass. This bias is entirely due to the fit to the deprojected density profile, hinting at the sensitivity of the deprojection procedure.

Among the widely adopted analytic profiles is the $\beta$-profile \citep{Cavaliere1976, Cavaliere1978}, employing the King approximation to the galaxy distribution profile arising from the isothermal model and assuming that the gas follows the mass distribution of galaxies within the cluster. The model has three free parameters: the normalisation, the power-law slope at large radii, and the turn-over radius between the core and power-law. This model describes a gas density distribution with a single power law and a core (i.e. a flat central density profile). In reality, observations find that many clusters have cusps (i.e. steep central density profiles), as well as a steepening of the slope at larger radii. To account for these two effects, the modified-$\beta$-model was introduced \citep{Vikhlinin2006_modbeta}, increasing the number of free parameters to seven. This model is no longer derived directly from the physics of an isothermal sphere since it includes empirical modifications, but it better represents the observed deprojected density profiles.

Alternatively, there is a range of methods for obtaining gas density profiles starting from a dark matter profile. For example, \citet{Sullivan2024} create a flexible profile which assumes distinct slopes for the inner and outer profiles, with an additional parameter controlling the transition. Similarly, \citet{Patej2015} create a family of models which transform a given dark matter density profile into a gas density profile, assuming an accretion shock is present in the gas. 

The observed temperatures in radial bins are similarly fit with an analytic profile. The prevailing choice is the 9-parameter model introduced by \citet{Vikhlinin2006_modbeta}. This model includes an overall normalisation, a term describing the temperature drop near the cluster centre, and a broken power law with a transition region describing the outer part of the cluster. The model's significant flexibility, arising from its numerous free parameters, enables accurate fitting of nearly all observed temperature profiles.

In this work, we will employ the FLAMINGO hydrodynamical cosmological simulations \citep{Schaye2023, Kugel2023} to study the hydrostatic mass bias of a large number of galaxy groups and clusters. In light of the limited number of massive galaxy clusters in previous hydrodynamical simulations, the unprecedented volume of the FLAMINGO simulations presents an important step forward. At $z=0$, FLAMINGO contains $>$$10^5$ clusters with $M_{\rm 500c}>10^{14} ~\mathrm{M_{\odot}}$ in their full cosmological context. Furthermore, the suite of simulations contains model variations varying the feedback strength, the type of AGN feedback, and the cosmology. The fiducial model has been shown to reproduce scaling relations and thermodynamic profiles from X-ray observations of galaxy clusters \citep{Braspenning2023}. In \citet{Kay2024}, we have shown that using Sunyaev-Zel'dovich effect weighted temperatures and focusing on the most massive clusters, the mass bias is only 5 per cent, which increases to 22 per cent when using X-ray weighted temperatures.

This paper is organised as follows, Section \ref{sec:methods} provides a description of the FLAMINGO simulations, the baryonic profiles and fitting procedure, and the computation of the hydrostatic mass. Section \ref{sec:results} presents our main results, Section \ref{sec:understand} aims to understand the origin of the bias and we summarise our findings in Section \ref{sec:conclusion}.

\section{Methods} \label{sec:methods}
In this section, we will describe the FLAMINGO simulations (\S \ref{sec:methods_simulations}), the construction of thermodynamic profiles (\S \ref{sec:methods:thermo_profiles}), the computation of hydrostatic masses (\S \ref{sec:methods_hsmass}), and the analytic profiles we fit to the simulated clusters (\S \ref{sec:analytic_profiles}). 
 
\subsection{The FLAMINGO simulations} \label{sec:methods_simulations}
FLAMINGO (Full-hydro Large-scale structure simulations with All-sky Mapping for the Interpretation of Next Generation Observations) is a large suite of hydrodynamical cosmological simulations, covering large cosmic volumes. The flagship run comprises a region of $\mathrm{(2.8 ~Gpc)^3}$, which is sufficiently large for statistical studies of galaxy groups and clusters. At $z=0$ FLAMINGO contains $>10^6$ haloes with $M_{\rm 500c} \geq 10^{13}~\mathrm{M_\odot}$, $>10^5$ haloes with $M_{\rm 500c} \geq 10^{14}~\mathrm{M_\odot}$, and $461$ haloes with $M_{\rm 500c} \geq 10^{15}~\mathrm{M_\odot}$. The simulations are described in detail in \citet{Schaye2023}. A unique feature is the machine learning-aided calibration of the subgrid prescriptions for stellar and AGN feedback to the observed $z=0$ galaxy stellar mass function and $z=0.1-0.3$ cluster gas fractions \citep{Kugel2023}. Variations on the fiducial model are made by shifting the observed gas fractions up and down by a multiple of their uncertainty ($\sigma$), and recalibrating the model to fit those new data points. FLAMINGO also includes models that vary the galaxy mass function or the cosmology. We will not use the latter here as we found that the cosmology variations do not affect the bias.

In this work we mainly use the redshift $z=0$ snapshot of the $\mathrm{(2.8 ~Gpc)^3}$ simulation, which has a gas particle mass $m_{\rm gas} = 1.07 \times 10^9 ~\mathrm{M_\odot}$, a dark matter particle mass $m_{\rm CDM} = 5.65 \times 10^9 ~\mathrm{M_\odot}$, a Plummer-equivalent comoving gravitational softening length $\epsilon_{\rm com} = 22.3 ~\mathrm{ckpc}$, and a maximum proper gravitational softening length $\epsilon_{\rm prop} = 5.70 ~\mathrm{pkpc}$. When comparing physics models in the FLAMINGO suite, we use the $\mathrm{(1 ~Gpc)^3}$ volumes, which were simulated using the same resolution as the $\mathrm{(2.8 ~Gpc)^3}$ run.

FLAMINGO uses the open source simulation code \textsc{swift} \citep{Schaller2023}, and solves the hydrodynamics using the \textsc{sphenix sph} scheme \citep{Borrow2022}. Massive neutrinos are included using the $\delta$f method \citep{Elbers2021}. The initial conditions are generated with a modified version of \textsc{monofonic} \citep{Hahn2021, Elbers2022}, and the cosmology `3x2pt + all external constraints' from the dark energy survey year 3 is used ($\Omega_{\rm m} = 0.306$, $\Omega_{\rm b} = 0.0486$, $\sigma_8 = 0.807$, $\mathrm{H_0} = 68.1~\mathrm{km~s^{-1}~Mpc^{-1}}$, $n_{\rm s} = 0.967$) \citep{Abbott2022}.

The FLAMINGO model includes subgrid implementations of radiative cooling \citep{ploeckinger2020}, star formation \citep{Schaye2008}, stellar mass loss \citep{Wiersma_stellarmassloss2009, Schaye2015}, supernova feedback \citep{DallaVecchia2008, Chaikin2022}, seeding and growth of supermassive black holes, and thermally-driven AGN feedback \citep{Springel2005, BoothSchaye2009, Bahe2022}. Two of the physics variations use kinetic jet feedback from AGN \citep{Husko2022}.

Haloes are found using a 3D Friends-of-Friends (FoF) algorithm, after which the center and substructure are identified using the 6D Friends-of-Friends (FoF) algorithm \texttt{VELOCIraptor} \citep{Elahi2019vr}. It defines the halo center as the particle with the lowest potential and separates bound particles of a single FoF group into a central and satellites. We then use the Spherical Overdensity and Aperture Processor (SOAP)\footnote{https://github.com/SWIFTSIM/SOAP} to compute halo properties within a range of apertures. The mass $M_{\rm 500c}$ used in this work is found by determining the radius at which the average internal density becomes 500 times the critical density of the universe, and summing the mass of all particles within that radius.

\subsection{Thermodynamic profiles} \label{sec:methods:thermo_profiles}
We make use of the temperature, density, and pressure profiles of massive objects ($M_{\rm 500c} > 10^{13}~\mathrm{M_\odot}$) in the FLAMINGO suite of simulations. The construction of these profiles, and their comparison to observations, is described in detail by \citet{Braspenning2023}. Here we provide a short overview.

Profiles are computed by dissecting a halo into 30 logarithmic 3D radial bins between 0.01 and 3.0 $r_{\rm 500c}$. Within each bin, a weighted average of the thermodynamic properties of the particles is computed. The average can be computed using different weights per particle. Unless stated otherwise, we use X-ray luminosity-weighted observer-frame [0.5-2.0 keV] profiles, because those are most relevant for the interpretation of and comparison with hydrostatic mass measurements. However, we will also compare with the bias inferred from volume weighted profiles, which gives better insight into the true dynamical state. To compute a weighted profile, each particle is multiplied by its value of the weighting quantity, then all weighted particles in the radial bin are summed, and finally, the sum is divided by the sum of the weights.

We exclude all particles which have experienced direct AGN heating in the last $15 ~\mathrm{Myr}$, as those might have spurious temperatures and densities. However, we find this to have no significant effect on the retrieved hydrostatic bias. Any particles in subhaloes are also removed, which is comparable to the removal of substructure in X-ray observations. In observations, only massive substructures that are distinct in 2D surface brightness maps can be removed. Our removal goes much further, removing all substructures, but we show in Appendix \ref{sec:subhaloes} that this makes no important difference.

Haloes are selected purely on their true mass, with no additional selection on their dynamical state. We checked for a correlation between the bias and relaxedness criteria, but did not find any (see Appendix \ref{sec:dynamical_state}).

We note that in \citet{Braspenning2023} we have shown that the thermodynamic profiles for the FLAMINGO simulations are generally in agreement with results from observational campaigns.

\subsection{Hydrostatic mass} \label{sec:methods_hsmass}

We define the hydrostatic mass bias as the difference from unity of the ratio of the hydrostatic mass to the true mass\footnote{Note that some authors define the hydrostatic bias as $1+b_{\rm HSE}$ or $-b_{\rm HSE}$.} measured directly from the particles in the simulation:
\begin{equation}
    b_{\rm HSE} = \frac{M_{\rm HSE, 500c}}{M_{\rm500c}} - 1\, .
\end{equation}
With this definition, an underestimate of the mass results in a negative bias.
The hydrostatic mass is defined as the mass of the halo inferred from the thermodynamic profiles under the assumption that it is in hydrostatic equilibrium. With the assumption of hydrostatic equilibrium (and in the absence of non-thermal pressure), the hydrostatic mass contained within a radius $r$ depends only on the derivatives of the temperature and density profiles, and the value of the temperature at that radius. This can be understood as a balance between the gravitational force and the thermal pressure gradient.

The hydrostatic mass can then be computed as
\begin{equation} \label{eq:hs_mass}
    M_{\rm HSE} (<r) = - \frac{k_{\rm B} T(r)}{G \mu m_{\rm p}} \left[ \frac{\mathrm{dlog}T(r)}{\mathrm{dlog}r} + \frac{\mathrm{dlog}\rho(r)}{\mathrm{dlog}r}\right] r \, ,
\end{equation}
with $k_\text{B}$ the Boltzmann constant, $G$ the gravitational constant,  $T(r)$ the temperature at radius $r$, $\mu = 0.59$ the mean particle mass of a fully ionised gas, $m_{\rm p}$ the proton mass, and $\rho(r)$ the density at radius $r$. If non-thermal pressure were accounted for, then there would be an additional term proportional to the spatial derivative of the non-thermal pressure.

In practice, when determining the hydrostatic mass $M_{\rm HSE, 500c}$, one first has to find the radius at which the internal density equals 500 times the critical density of the universe, $\rho_{\rm 500c}$. We find this radius by minimizing the following quantity
\begin{equation}
    \delta M = \left| \frac{4}{3} \pi r^3 \rho_{\rm 500c} - M_{\rm HSE}(r) \right| \, ,
\end{equation}
where we denote the resultant radius $r_{\rm 500c, HSE}$. The hydrostatic mass can then be computed using 
\begin{equation} \label{eq:hs_mass_2}
    M_{\rm HSE, 500c} = \frac{4}{3} \pi r_{\rm 500c, HSE}^3 \rho_{\rm 500c} \, .
\end{equation}

\begin{figure*}
    \centering
    \includegraphics[width = \linewidth]{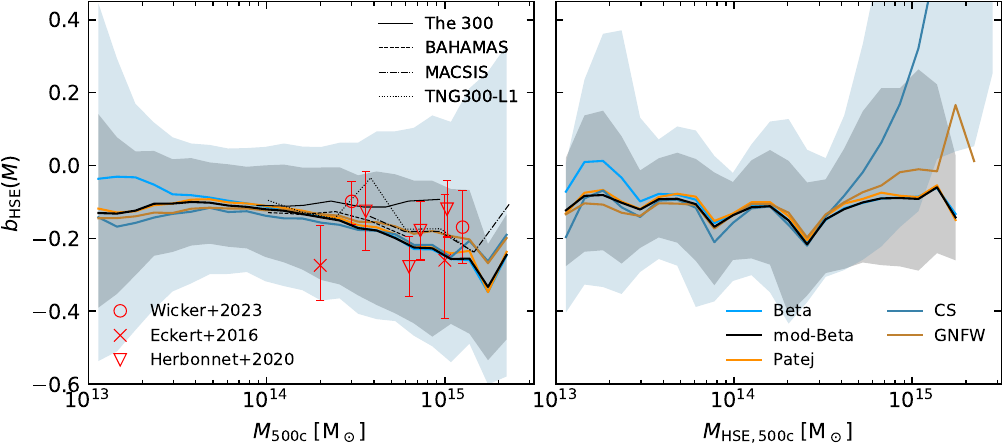}
    \caption{The median hydrostatic mass bias, calculated using different methods, as a function of the true mass (left panel) or the inferred hydrostatic mass (right panel). Different colours indicate the use of different analytic profiles to estimate the derivatives of the density and temperature profiles at $r_{\rm 500c}$. The shaded areas show the $16^{\rm th}$-$84^{\rm th}$ percentile ranges for the modified-$\beta$ and cubic-spline methods. Thin black lines in the left panel show predictions for massive objects from other simulations. Red data points with error bars are observational measurements from comparisons between X-ray and weak lensing masses. The predictions are in good agreement with the data.}
    \label{fig:fit_options_r500hs}
\end{figure*}

\subsection{Fitting the thermodynamic profiles} \label{sec:analytic_profiles}

Following the approach of observational studies \citep[e.g.][]{Turner2024, Eckert2016, Vikhlinin2006}, we fit our 3D density and temperature profiles with analytic functions, which then provide smooth, well-defined derivatives at any radius. To compute the hydrostatic mass, only the profile around $r_{\rm 500c}$ is important, hence we limit our fits to the region 0.5-2.0 $r_{\rm 500c}$.

For the gas density profile, we make use of three different analytic profiles. The first is the 3-parameter $\beta$-profile,
\begin{equation}
    \rho_{\rm g}(r) = \rho(0) \left( 1 + \left[\frac{r}{r_{\rm c}}\right]^2\right)^{-3\beta / 2} \, ,
\end{equation}
where the central density, $\rho(0)$, the core radius, $r_{\rm c}$, and $\beta$, which determines the slope at large radii, are free parameters. 

The second is the modified-$\beta$-profile, which is our fiducial choice,
\begin{equation}
    \rho_{\rm g}(r) = \left[ \rho(0)^2 \frac{\left[\frac{r}{r_{\rm c}}\right]^{-\alpha}}{\left(1 + \left[\frac{r}{r_{\rm c}}\right]^2\right)^{3\beta - \alpha/2}} \left( \frac{1}{1 + \left[\frac{r}{r_{\rm s}}\right]^\gamma} \right)^{\varepsilon / \gamma} \right]^{1/2} \, ,
\end{equation}
where $\alpha$, $\beta$, $\gamma$, $\epsilon$ are free parameters, as well as the normalisation of the density profile $\rho(0)$, the core radius $r_{\rm c}$, and the scale radius $r_{\rm s}$. The scale radius ($r_{\rm s}$) allows for a second characteristic transition radius in the profile. Though this model has 7 free parameters, $\gamma$ is conventionally set to $\gamma = 3$, which we adopt throughout this work. Note that because we only fit the region 0.5-2.0 $r_{\rm 500c}$, the parameters related to the core remain largely unconstrained.

Though the modified-$\beta$-profile tends to be a better fit to galaxy cluster profiles, the modifications are empirical and lack a theoretical footing. An alternative to the approach of adding modifications to the simple $\beta$-profile, are the family of gas density profiles introduced by \citet{Patej2015}. Starting from spherically symmetric dark matter density profiles, they obtain analytic expressions for gas density profiles under minimal assumptions, apart from requiring a virial shock to be present. The profile we utilise is the gas density profile in that family belonging to an NFW dark matter profile. This is our third analytic function for the density profile,

\begin{equation}
    \rho_{\rm g}(r) = \Gamma A \frac{(r/s)^{2\Gamma - 2}}{r/r_{\rm s} \left( 1 + (s / r_{\rm s}) (r/s)^\Gamma \right)^2} \, ,
\end{equation}
with $r_{\rm s}$ the NFW scale radius, $s$ the radius where the virial shock occurs, $\Gamma$ the ratio of the gas to dark matter shock jump parameters, and $A$ a normalisation. Hence there are four free parameters.

Instead of fitting the density and temperature profiles separately, some authors choose to fit the pressure profile \citep[e.g.][]{Kay2024, Gupta2017, Planelles2017, Pearce2020, Nagai2007}, which is the product of the two. The equation for the hydrostatic mass (Eq.~\ref{eq:hs_mass}) then simplifies to only requiring the pressure derivative. Here we will fit pressure profiles with a generalised NFW profile (GNFW),
\begin{equation}
    P_{\rm g}(r) = \frac{P_0}{(c_{\rm 500c} r)^\gamma \left( 1 + (c_{\rm 500c}r)^\alpha \right)^{(\beta - \gamma)/\alpha}} \, ,
\end{equation}
where we treat $P_0$, $c_{\rm 500c}$, $\alpha$, and $\beta$ as free parameters, and we use $\gamma = 0.31$ in line with \citet{Arnaud2010, Barnes2017}.

The temperature profile is always fit using the universal temperature profile introduced by \citet{Vikhlinin2006_modbeta},
\begin{equation}
    T(r) = T_{\rm 0} \left[ \frac{\left[\frac{r}{r_{\rm c}}\right]^{\alpha_{\rm c}} + \frac{T_{\rm min}}{T_{\rm 0}}}{ 1 + \left[\frac{r}{r_{\rm c}}\right]^{\alpha_{\rm c}}} \right] \left[ \frac{\left[\frac{r}{r_{\rm t}}\right]^{-\alpha}}{\left(1 + \left[\frac{r}{r_{\rm t}}\right]^{\beta} \right)^{\gamma / \beta}} \right] \, ,
    \label{eq:temperature_analytic}
\end{equation} 
an eight-parameter function, where $T_{\rm 0}$ represents the temperature normalisation, $T_{\rm min}$ the minimum temperature of the core, $r_{\rm c}$ the core radius, $r_{\rm t}$ the transition radius, which, along with $\alpha$, $\alpha_{\rm c}$, $\beta$, and $\gamma$, are free parameters. Similar to the modified-$\beta$-profile, out choice to fit between 0.5-2.0 $r_{\rm 500c}$ leaves the parameters related to the core largely unconstrained.

These functions are fit to the thermodynamic profiles using a least squares minimization process in logarithmic space under the assumption of uniform errors, employing the \texttt{Nelder-Mead} algorithm \citep{Gao2012}.

Finally, as an alternative to fitting analytic profiles, we employ non-parametric cubic-spline interpolation between the data points of the raw profiles computed from the simulations. Applying a cubic-spline interpolator then gives access to smooth values and derivatives at any given radius of the profile. When using the cubic-spline in this work, we will apply it to both the density and temperature profiles. For all other methods, the temperature profile is fit using Eq.~\ref{eq:temperature_analytic}.

\section{Results} \label{sec:results}
In this section we will investigate the dependence of the hydrostatic bias on mass (\S\ref{sec:mass}), the choice of analytic profile (\S\ref{sec:analytic_profile}), the weighting scheme for the thermodynamic profiles (X-ray flux versus volume weighted averaging; \S\ref{sec:weighting}), the effect of physics model variations within FLAMINGO on the bias (\S\ref{sec:model_variations}), and the dependence on redshift (\S\ref{sec:redshift}).

\subsection{Dependence on mass} \label{sec:mass}

The black solid line in Fig.~\ref{fig:fit_options_r500hs} shows the median hydrostatic mass bias predicted by our fiducial model (the $z=0$ output from the 2.8~Gpc simulation, X-ray luminosity weighted thermodynamic profiles fit with a modified-$\beta$ density profile and a \citet{Vikhlinin2006} universal temperature profile). The left and right panels show the hydrostatic bias as a function of the the true mass ($M_{\rm 500c}$) and the hydrostatic mass ($M_{\rm HSE, 500c}$), respectively. The shaded regions indicate the 16$^{\rm th}$-84$^{\rm th}$ percentiles for the modified-$\beta$-profile and cubic-spline interpolation. 
The scatter is smallest for true masses $\sim 10^{14}\,\mathrm{M_\odot}$. 

When binning in true mass all methods exhibit a significant increase in the median absolute bias in the massive cluster regime, signalling a substantial underestimation of cluster mass. This underestimation increases to an offset of 25 per cent for the most massive objects. The cubic-spline and GNFW methods have slightly smaller bias compared to the others. However, when binning in inferred hydrostatic mass, the bias is almost mass-independent for all parametric methods, hovering around a value of $b_{\rm HSE} \approx -0.1$. 

Within the cluster regime ($M_{\rm 500c} ~/~\mathrm{M_\odot} > 10^{14.5}$), the results thus depend strongly on whether we bin in true or inferred hydrostatic mass. Since there are relatively few high-mass objects, up-scatter through a positive bias from lower masses can come to dominate the highest-mass bins, raising the median bias above zero, when expressed as a function of $M_{\rm HSE, 500c}$. The cubic-spline interpolation method is especially susceptible to this effect, which can be understood due to the increase in scatter at high true mass as seen from the shaded blue region in the left panel.

Our results for the median X-ray weighted bias as a function of true mass align closely with the results from \citet{Barnes2021} utilising mock X-ray observations on BAHAMAS, MACSIS, and IllustrisTNG as shown by thin broken black lines in Fig.~\ref{fig:fit_options_r500hs}. The bias for the 300 simulations \citep{Gianfagna2023}, shown by the solid black line, is significantly less mass-dependent and only agrees with FLAMINGO around $10^{14} ~\mathrm{M_\odot}$. We note that the 300 simulations were fit using mass-weighted profiles.

Observational determinations of the bias, shown as the red data points, scatter widely below $10^{15}~\mathrm{M_\odot}$. We extract the data points from the observational literature by taking the highest and lowest mass in each observed sample and applying the power-laws, with errors, fitted by the respective authors at those masses. These data points are obtained using either $\beta$- or double-$\beta$-profiles to fit the density profile, and the \citet{Vikhlinin2006} prescription for the temperature profile. Our results agree very well with \citet{Wicker2023} at $3\times 10^{14}~\mathrm{M_\odot}$, and have a very similar mass trend. In the study from \citet{Eckert2016} the inverse mass trend is found, where the bias decreases with increasing mass, though the magnitude is very similar to our result and the trend is weak. \citet{Herbonnet2020} find no clear mass trend with a slight dip in the bias around $10^{14.8}~\mathrm{M_\odot}$. We conclude that the FLAMINGO hydrostatic biases are in good agreement with the observations, but that the observational constraints are rather uncertain for all but the highest masses.

We note that during the calibration of the FLAMINGO subgrid feedback to the observed cluster gas fractions, a constant hydrostatic mass bias of -0.257 was assumed for gas fractions inferred from X-ray observations \citep{Kugel2023}. This value was taken from weak lensing calibrations and agrees well with the FLAMINGO predictions for the most massive clusters. In appendix~\ref{sec:recalibration} we show that recalibrating the FLAMINGO model to data corrected using the hydrostatic mass-dependent bias predicted here only leads to small differences in the gas fractions. The exception is the group regime, where the gas fractions become significantly higher, but where there is no lensing data to compare the predicted bias with.

\subsection{Dependence on the analytic profile} \label{sec:analytic_profile}

We explore how the choice of analytic profile affects the hydrostatic mass, and consequently, the hydrostatic bias predicted by the simulations. The hydrostatic bias at $r_{\rm HSE, 500c}$ is computed using the five distinct methods introduced in section \ref{sec:analytic_profiles}:
\begin{itemize}
    \item[1.] $\beta$-profile
    \item[2.] modified-$\beta$-profile
    \item[3.] Patej-profile
    \item[4.] GNFW-profile
    \item[5.] cubic-spline interpolation
\end{itemize}
The results are depicted as different coloured lines in Fig.~\ref{fig:fit_options_r500hs}. The scatter is only shown for the fiducial modified-$\beta$ and cubic spline profiles. For all parametric methods the scatter is similar to that for the fiducial profile, but the cubic spine method yields a larger scatter.

Within the galaxy group regime ($10^{13.0} < M_{\rm 500c} ~/~\mathrm{M_\odot} < 10^{13.5}$), the $\beta$-profile method exhibits a significantly smaller median bias (i.e.\ a smaller absolute value of $b$) than the other methods. 
For objects straddling the boundary between groups and clusters ($10^{13.75} < M_{\rm 500c} ~/~\mathrm{M_\odot} < 10^{14.25}$), the five methods converge to a median bias of approximately $b_{\rm HSE} \approx -0.1$. These findings hold both when binning in true and in hydrostatic mass.

For the remainder of this work, we will employ the modified-$\beta$-profile as a middle-of-the-road model. We note that none of our conclusions change if we opt for any of the other parametric profiles.

\subsection{X-ray versus volume weighted averaging} \label{sec:weighting}

Because hydrostatic masses are inferred from spherically averaged profiles, any deviations from spherical symmetry will complicate their interpretation. The presence of structure will lead to deviations between thermodynamic profiles computed using different weighting schemes. Our approach of using X-ray weighted thermodynamic profiles in the hydrostatic mass inference is appropriate when comparing with results from X-ray observations, but to measure the true dynamical state of the cluster volume weighting is the natural choice.  If the gas is clumpy, then X-ray weighting will favour denser gas, provided it is not sufficiently dense to cool to low temperatures, because the emissivity scales with the square of the density. In Appendix \ref{sec:subhaloes} we study the impact of subhaloes on the hydrostatic bias, and find it to be relatively small.

\begin{figure}
    \centering
    \includegraphics[width=\linewidth]{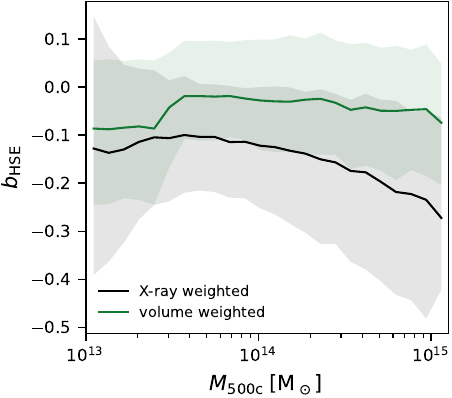}
    \caption{The median hydrostatic mass bias, calculated using the modified-$\beta$ method from X-ray weighted (black) and volume weighted (green) thermodynamic profiles, as a function of the true mass. Shaded areas show the 16th-84th percentile ranges. The volume weighted bias is significantly smaller and less mass dependent compared to the X-ray weighted hydrostatic mass bias.}
    \label{fig:hs_bias_weight_comparison}
\end{figure}

We compare the hydrostatic mass bias measured from X-ray-  and volume weighted profiles. In earlier work we compared with mass-weighted profiles, and found them to fall in-between the X-ray- and volume-weighted cases \citep{Braspenning2023}. From Fig.~\ref{fig:hs_bias_weight_comparison}, it is clear that the X-ray-weighted profiles result in a significantly stronger, and more mass dependent bias. As a large mass bias is driven by shallow gradients, this indicates that the thermal pressure gradient in X-ray-weighted gas is shallower. We conclude that for $M_\text{500c} \gg 10^{13}\,\text{M}_\odot$ most of the hydrostatic mass bias is due to X-rays being a biased tracer of the gas rather than due to true deviations from hydrostatic equilibrium.

\subsection{Model variations} \label{sec:model_variations}

\begin{figure*}
    \centering
    \includegraphics[width = \linewidth]{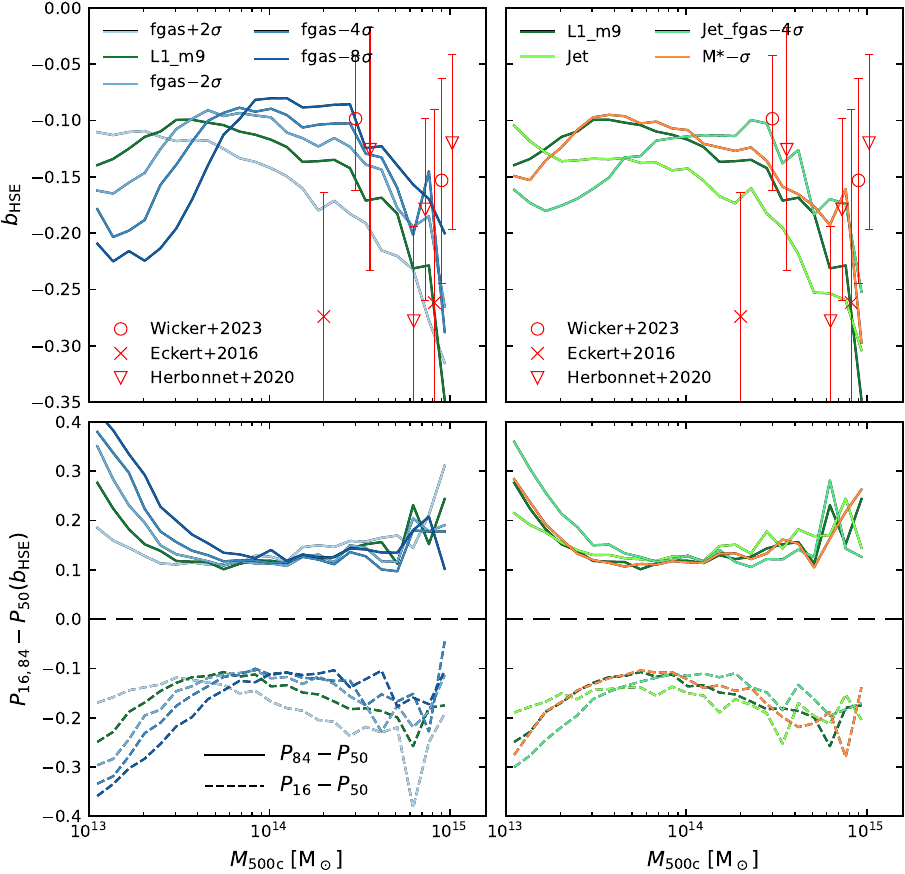}
    \caption{Top panels show the median hydrostatic mass bias as a function of true mass for different FLAMINGO physics variations (different colours). Bottom panels show the difference between the $16^{\rm th}$ percentiles (dashed), and $84^{\rm th}$ (solid) percentiles and the median of each model variation. There are large differences in both the median bias and scatter between models. Red scatter points indicate observed bias values, with the highest mass taken to be $10^{14.9}~\mathrm{M}_\odot$ to align with the mass range probed by the L1\_m9.}
    \label{fig:variations}
\end{figure*}

\begin{figure*}
    \centering
    \includegraphics[width = \linewidth]{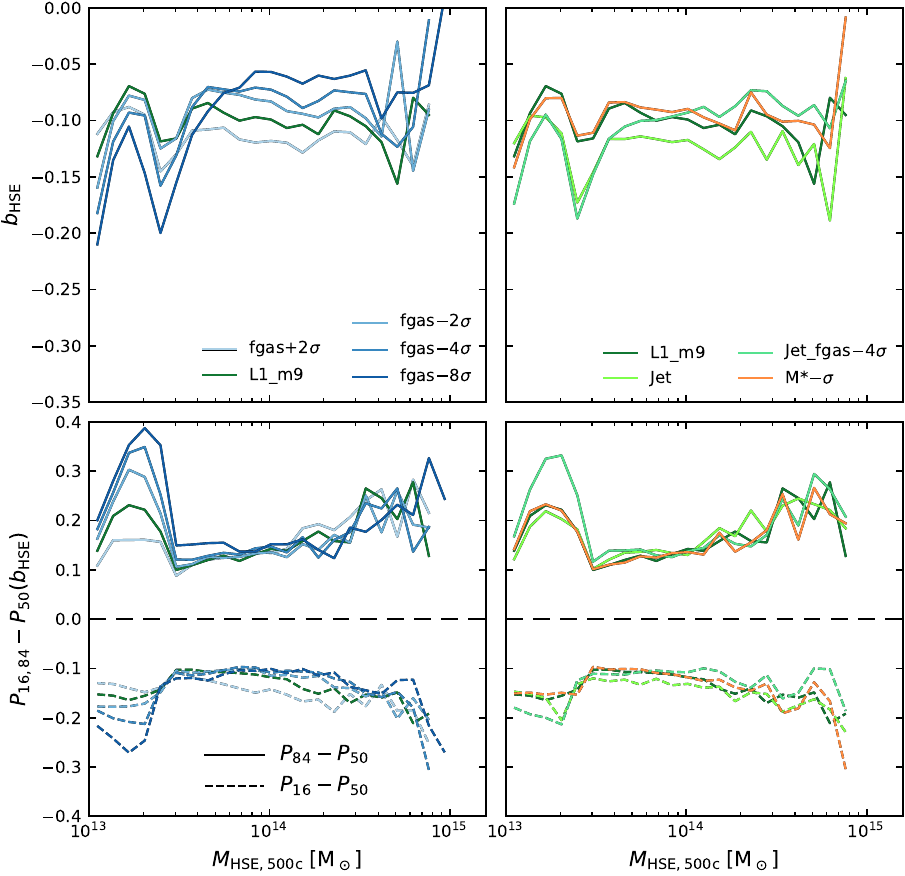}
    \caption{As Fig.~\ref{fig:variations} but for the bias as a function of the inferred hydrostatic mass $M_{\rm HSE, 500c}$. Compared to the bias as a function of true mass shown in Fig.~\ref{fig:variations}, the median bias curves are less mass dependent, but the magnitude of the offset between the different physics models is comparable. The scatter is less sensitive to the physics model than in Fig.~\ref{fig:variations}.}
    \label{fig:variations_hse}
\end{figure*}

A key feature of the FLAMINGO suite of simulations is the availability of simulations that vary the cluster gas fractions and/or stellar mass function, and the AGN feedback mechanism. Different gas fractions in massive haloes yield different thermodynamic profiles. Particularly notable are the Jet models of AGN feedback, which exhibit a distinctly different distribution of gas, especially in the cluster cores \citep[see][]{Braspenning2023}. Because those differences are largest in cluster cores, they might have only a small effect on the bias which is measured around the overdensity radius $r_{\rm 500c}$. Fig.~\ref{fig:variations} illustrates different bias values obtained for the different models as a function of the true mass, employing the modified-$\beta$-profile. The top panels show the median values (note the narrower y-axis range compared to Fig.~\ref{fig:fit_options_r500hs}), and the bottom panels the difference between the 84$^{\mathrm{th}}$ percentile and the median (solid lines) and between the 16$^{\mathrm{th}}$ percentile and the median (dashed lines).

\begin{figure*}
    \centering
    \includegraphics[width = \linewidth]{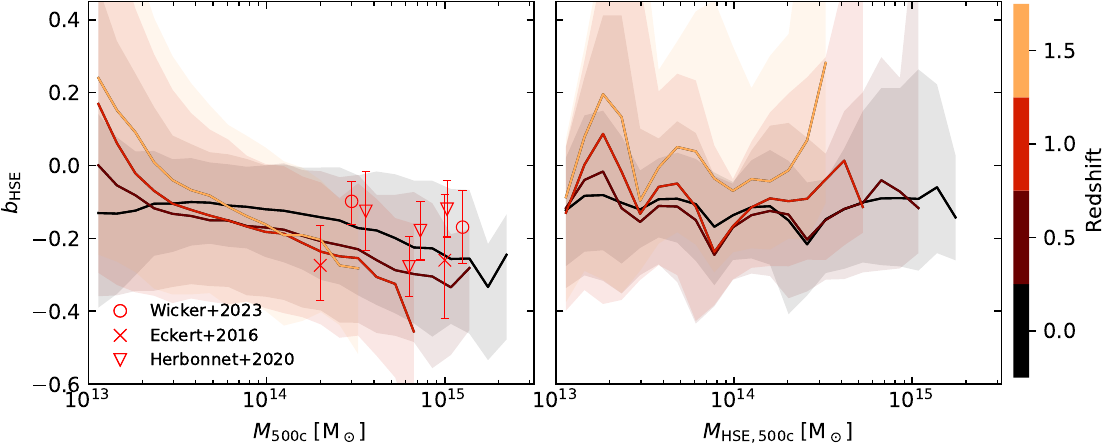}
    \caption{As Fig.~\ref{fig:fit_options_r500hs}, but showing the bias inferred using the modified-$\beta$-profile at different redshifts (different colours). As a function of true mass, the bias curves steepen with increasing redshift. However, at a fixed inferred hydrostatic mass they are almost unchanged except at the highest redshift considered ($z=1.5$). The observational data points are all for clusters with $z \leq 0.5$}
    \label{fig:redshift}
\end{figure*}

In the top left panel, we observe a dip where the bias becomes strongly negative at low masses, after which the bias recovers to $b_{\rm HSE} = -0.1$ before again becoming increasingly negative toward higher masses. The depth and mass at which the dip occurs increase with decreasing gas fraction, from having no such dip for fgas$+2\sigma$, to $b_{\mathrm{HSE}} \approx -0.15$ at $10^{13} ~\mathrm{M_\odot}$ for fgas$-2\sigma$ and increasing towards $b_{\mathrm{HSE}} \approx -0.22$ at $2 \times 10^{13} ~\mathrm{M_\odot}$ for fgas$-8\sigma$. Apart from the horizontal shift, the shape of the bias curve seems identical for all gas fraction variations. The difference in median bias at a given mass between the two most extreme models $b_{\mathrm{HSE}} [{\text{fgas}-8\sigma}]$ and $ b_{\mathrm{HSE}} [{\text{fgas}+2\sigma}]$, is around 0.1 for most masses, only around the turnover point at $4 \times 10^{13} ~ \mathrm{M_\odot}$ is the difference significantly smaller.

We note that between $7\times10^{14}~\mathrm{M_\odot}$ and $10^{15}~\mathrm{M_\odot}$, the bias inferred from observations shows significant scatter. It ranges from a bias exceeding that found for the highest gas fraction variation in FLAMINGO \citep[][at $7\times10^{14}~\mathrm{M_\odot}$]{Herbonnet2020} to a smaller bias than predicted by the lowest gas fraction variation \citep[][at $10^{15}~\mathrm{M_\odot}$]{Wicker2023, Herbonnet2020}. In general, the difference between the gas fraction variations is of the same order as the error on the observationally determined bias.

The top right panel shows that the model with a reduced galaxy stellar mass function predicts a hydrostatic bias that is very similar to that of the fiducial model, with just a slight offset in the same direction as decreasing gas fraction. That is unlike the jet model, which is almost identical to the fgas$+2\sigma$ model, at all masses. At low masses, this can be explained due to the increased gas fraction in that model for groups \citep[see Figure 10 of][]{Schaye2023}, however, for clusters the gas fractions are identical to the fiducial model. The change in AGN feedback mode seemingly affects the gas distribution out to large radii, yielding temperature and density gradients similar to the increased gas fraction model.
The Jet\_fgas$-4\sigma$ model exhibits a horizontal offset close to that of the fgas$-4\sigma$ model, but the dip in the bias at low mass reaches a minimum value of $b_{\mathrm{HSE}} \approx -0.17$ compared with $b_{\mathrm{HSE}} \approx -0.20$ for fgas$-4\sigma$, yielding a flatter bias curve with mass. The flattening of the bias curve only occurs below $10^{14} ~\mathrm{M_\odot}$, indicating that the change in AGN feedback mode has its largest effect on group mass haloes.

The scatter shown in the lower panels has a similar horizontal shift between the models with different gas fractions as for the median curves in the upper panels. The minimum spread of $\sigma \approx 0.2$ is reached at $3\times 10^{13} ~\mathrm{M_\odot}$ for fgas$+2\sigma$ after which the scatter gradually increases towards higher masses reaching $\sigma \approx 0.35$ at $5\times 10^{14}~\mathrm{M_\odot}$, whereas for the lowest gas fraction model the minimum spread is reached at $10^{14} ~\mathrm{M_\odot}$, resulting in a smaller spread of $\sigma \approx 0.25$ at $5\times 10^{14}~\mathrm{M_\odot}$.

The increase of scatter towards larger masses is dominated by the downscatter which increases faster than the upscatter, indicating that the distribution becomes skewed at high masses. Below the mass of minimum spread, the scatter in the bias diverges, with the lower gas fraction model variations having more scatter. 

The bottom right panel shows that the stellar mass variation ($M_\star - \sigma$) is very close to the fiducial model (L1\_m9), but slightly offset in the direction of fgas$-2\sigma$. The jet model is similar to fgas$+2\sigma$, and the Jet\_fgas$-4\sigma$ model is as fgas$-4\sigma$ but suppressed below $10^{14}~\mathrm{M_\odot}$.

It should be noted that although the differences in bias between the models are large at all masses, they remain within the bounds set by the $1\sigma$ percentiles of the fiducial model.

Fig.~\ref{fig:variations_hse} shows the same result but as a function of the inferred hydrostatic mass, which is the mass that would be measured in observations where the true mass cannot be accessed (i.e. X-ray observations). The top panels indicate that the offsets in the median bias values persist, though the mass of the crossing point is now the same for all models at $M_{\rm HSE, 500c} \approx 4 \times 10^{14} ~\mathrm{M_\odot}$. After this crossing point, the bias curves flatten instead of decreasing with increasing mass. They settle around a value of $b_{\mathrm{HSE}} \approx -0.1$, lower gas fractions yielding a smaller $b_{\rm HSE}$, i.e. a more biased measurement. The jet and stellar mass model variations have the same relative behaviour to the gas fraction variations as when comparing with the true mass $M_{\mathrm{500c}}$.

The scatter (bottom panels) shows a slightly increasing trend with increasing mass. For $M_{\rm HSE, 500c} > 3 \times 10^{13} ~ \mathrm{M_\odot}$ the scatter is relatively insensitive to the physics model.

\subsection{Redshift evolution} \label{sec:redshift}

The left panel of Fig.~\ref{fig:redshift} shows that the most massive objects at each redshift have a median bias of $b_{\rm HSE} \approx -0.3$. The bias is a decreasing function of true mass at all redshifts. The most striking feature at $z > 0$ is a sharp upturn of the hydrostatic mass for objects between $1-3 \times 10^{13}$ with increasing redshift. The mass of those objects is overestimated by $\approx 20$ per cent at $z > 1$. A positive bias can be explained by either an enhanced temperature or an increase in the density and temperature gradients. This seems to occur preferentially for lower mass objects at higher redshifts.

Generally, the median bias curve for $M_{\rm 500c}$ steepens with increasing redshift, making the bias more strongly mass dependent. The right panel shows that as a function of the inferred hydrostatic mass ($M_{\rm HSE, 500c}$) the bias remains insensitive to mass as the redshift increases but weakens for $z\gtrsim 1$.
The scatter in the bias increases with redshift. 

Between $z = 0$ and $z = 1.5$ the number of haloes decreases from $>$$10^6$ to $>$$10^5$ for $M_{\rm 500c} \geq 10^{13}~\mathrm{M_\odot}$, from $>$$10^5$ to $>$$10^3$ for $M_{\rm 500c} \geq 10^{14}~\mathrm{M_\odot}$, and from $461$ to $0$ for $M_{\rm 500c} \geq 10^{15}~\mathrm{M_\odot}$.

\section{Origin of the mass bias} \label{sec:understand}
\begin{figure*}
    \centering
    \includegraphics[width = \linewidth]{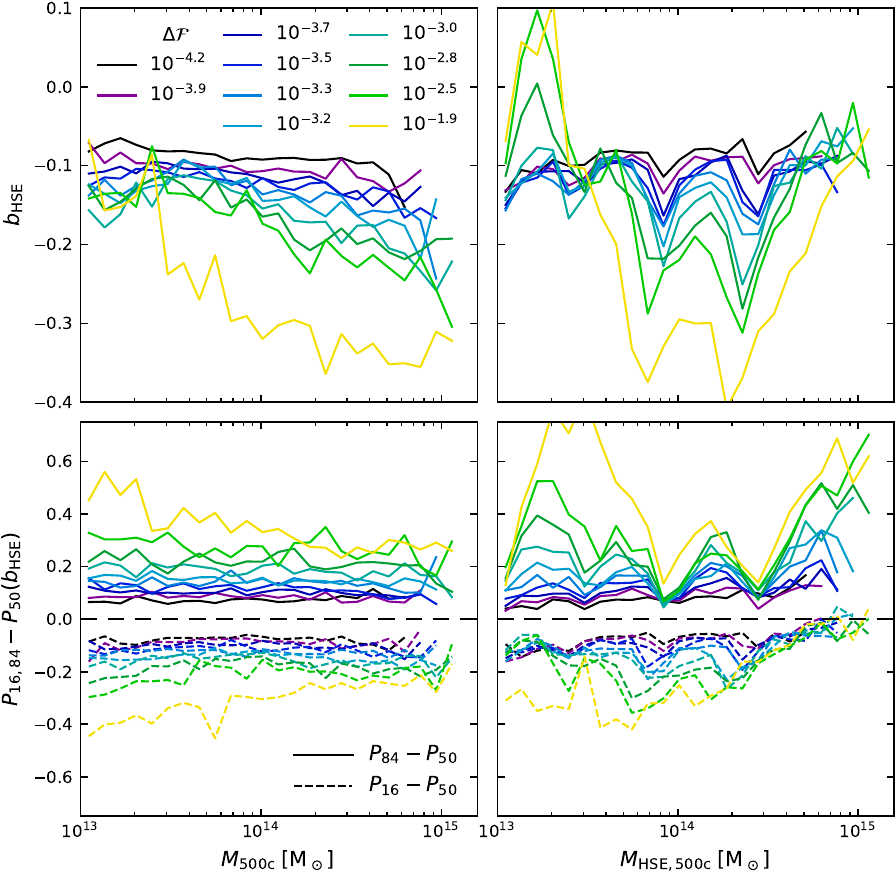}
    \caption{As Fig.~\ref{fig:variations}, but showing the hydrostatic mass bias inferred using the modified-$\beta$ fit to the density profile in bins of  $\Delta \mathcal{F}$, the root mean square error of the fitted density and temperature profiles. Each line corresponds to a subset of 10 per cent of all haloes, with the colour indicating the median $\Delta \mathcal{F}$ in that subset (a lower $\Delta \mathcal{F}$ indicates a better fit). Poorer fits tend to correspond to a larger absolute value of, and more scatter in the bias. The effect is stronger at larger true masses.}
    \label{fig:gof_biases}
\end{figure*}
This section will investigate the origin of the bias. We first study the effect of the fit quality on the obtained bias in Section \ref{sec:fit_quality}, followed by a characterisation of the objects with outlying bias values for their mass in Section \ref{sec:scatter}. We then study the contribution of the constituent components of the hydrostatic mass equation to the bias in Section \ref{sec:origin_of_bias}. Finally, we study the energy balance of virial equilibrium in Section \ref{sec:energy_balance}. 

\subsection{Fit quality} \label{sec:fit_quality}

The stronger hydrostatic mass bias at high true masses (Fig.~\ref{fig:fit_options_r500hs}), combined with the enhanced scatter for the bias values at high mass inferred using the cubic-spline method to fit the profiles, prompts us to investigate the fit quality across the entire mass range. To this end, we compute a goodness-of-fit (GOF) statistic for each fit to the density or temperature profile (using the modified-$\beta$-profile to fit the density profile),
\begin{equation}
    \Delta \mathcal{F} = \sqrt{\frac{1}{N_{\rm bins}} \sum_{i = 1}^{N_{\rm bins}} \left[ \log_{10} \left( \mathcal{F} / \mathcal{F}_{\rm analytic} \right) \right]^2} \, ,
\end{equation}
which is the root mean square difference between the logarithm of the profile $\mathcal{F}$ and the best fit, $\mathcal{F_{\rm analytic}}$; $N_{\rm bins} = 8$ is the number of radial bins in the profile between $0.5~r_{\rm 500c}$ and $2.0~r_{\rm 500c}$.

\begin{figure*}
    \centering
    \includegraphics[width = \linewidth]{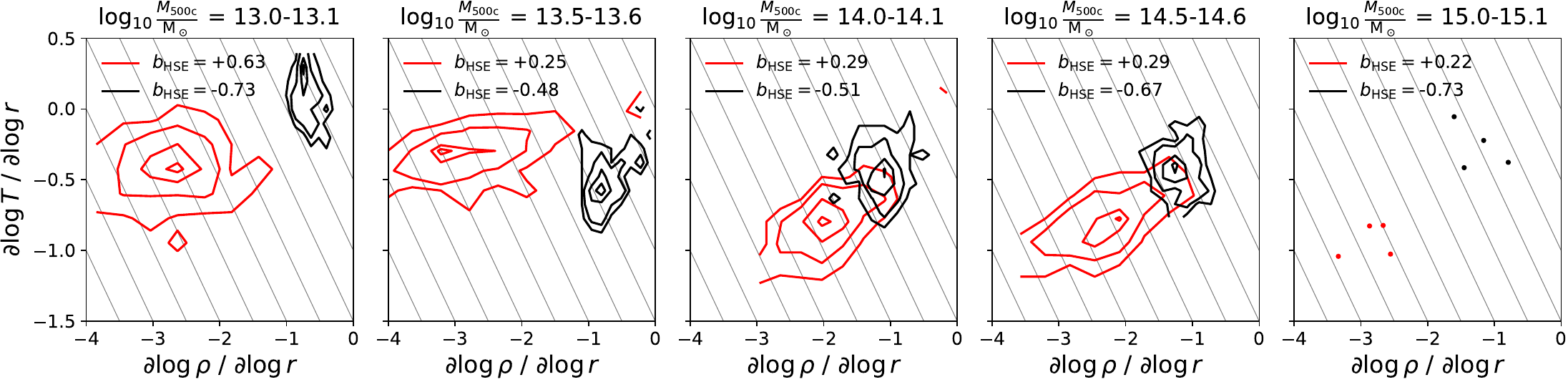}
    \caption{The logarithmic derivatives of the luminosity-weighted modified-$\beta$ density profiles and the luminosity-weighted temperature profiles, at $r_{\rm HSE,500c}$, for the objects with the 5 per cent lowest hydrostatic masses (black) or 5 per cent highest masses (red) in 5 narrow true mass bins. The legend in each panel indicates the median bias values of each subset. The two sets of contours are consistently offset and can be connected by moving almost orthogonal to the thin black diagonal lines which correspond to a constant pressure derivative, showing that the difference in bias could be explained by differences in the profile derivatives at a constant temperature (see Eq.~\ref{eq:hs_mass}).}
    \label{fig:outliers_derivs_modbeta}
\end{figure*}

Fig.~\ref{fig:gof_biases} shows how the bias changes in 10 bins of decreasing fit quality. Each bin contains 10 per cent of all objects, and the legend denotes the median combined $\Delta \mathcal{F}$ from both the density and temperature profile in that bin ($\Delta \mathcal{F}_\rho + \Delta \mathcal{F}_T$). The median bias increases toward larger negative values of $b_{\rm HSE}$ with decreasing fit quality (i.e. increasing $\Delta \mathcal{F}$). 

As a function of the true mass $M_{\rm 500c}$, apart from increasing in magnitude, the bias also becomes more mass dependent for poorer fits. The quality of the fit is particularly important at high masses. For the 30 per cent of massive clusters ($M_{\rm 500c} \sim 10^{15} ~ \mathrm{M_\odot}$) with the poorest fits, the bias is close to that inferred from observations. 

As a function of the inferred hydrostatic mass $M_{\rm HSE, 500c}$, poorer fits incur a larger absolute bias in the mass range $4 \times 10^{13} - 5 \times 10^{14} ~\mathrm{M_\odot}$, whereas they converge towards $b_{\rm HSE} \approx 0.1$ for larger masses. 

The bottom panels of Fig.~\ref{fig:gof_biases} show that the scatter increases with increasing $\Delta \mathcal{F}$. For the worst fits (yellow lines) the spread in bias values reaches $\sigma = 0.6$ for $10^{14}~\mathrm{M_\odot}$ haloes.   

In the bottom right panel, we observe a highly asymmetric scatter for $M_{\rm HSE, 500c} > 5 \times 10^{14} ~ \mathrm{M_\odot}$, for all values of  $\Delta \mathcal{F}$ the upscatter is significantly larger than the downscatter. At the highest masses ($M_{\rm HSE, 500c} \sim 10^{15}~\mathrm{M_\odot}$) the downscatter approaches zero, and the bias is positive for nearly all haloes. 

Both as a function of true mass and inferred hydrostatic mass, the best 20 per cent of fits yield an almost constant bias ($b_{\rm HSE} \approx -0.1)$, with small scatter, $\sigma_{\rm HSE} = 0.1$. 

Intuitively, these results make sense as more massive objects are more likely to be disturbed and hence have fluctuations in their density and temperature profiles, particularly in the outskirts \citep[see also][]{Kay2024}. The fluctuations are always positive, since unmasked, intervening structure results in a local increase in the density field, and the luminosity-weighted density profiles will be higher if the gas is more clumpy. When fitting the binned profiles, such bins with increased densities tend to flatten the profile, yielding smaller derivatives, smaller hydrostatic masses (see Eq.~\ref{eq:hs_mass}) and hence larger negative biases.

We note that even the 10 per cent worst fits (yellow line), still have small errors, of order $0.1~\mathrm{dex}$ per data point. We checked that the other analytic fits to the density profile yield similar results.

\subsection{Drivers of outlying hydrostatic mass} \label{sec:scatter}
This section investigates the scatter in hydrostatic mass bias at fixed mass by identifying the objects with the 5 per cent highest and lowest bias values within each true mass bin. To ensure that we do not see any residual dependence on the goodness-of-fit (as shown in Fig.~\ref{fig:gof_biases}), we match the $\Delta \mathcal{F}$ distributions of the two subsamples containing outliers. About $70$ per cent of the objects in the subsamples can be matched, for each object in a sample the other sample has an object for which the $\Delta \mathcal{F}$ differs by less than 10 per cent. Leveraging the mathematical expression for hydrostatic mass (Eq.~\ref{eq:hs_mass}), we analyze the derivatives with radius of the temperature and density profiles for these two groups of objects. 

\begin{figure}
    \centering
    \includegraphics[width = \linewidth]{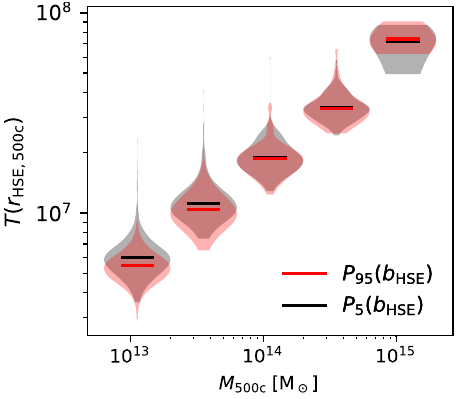}
    \caption{The luminosity-weighted temperature at $r_{\rm HSE, 500c}$ of objects with the 5 per cent lowest hydrostatic masses, i.e. the $5^{\rm th}$ percentile $P_5$, (black) or 5 per cent highest hydrostatic masses, i.e. the $95^{\rm th}$ percentile $P_{95}$, (red) in 5 different bins of true mass. The medians (horizontal solid lines) and distributions of temperatures (the width of the shaded region is proportional to the number of objects) are comparable between the two samples for all mass bins. This indicates that temperature differences cannot explain the different hydrostatic masses.}
    \label{fig:outliers_temp_modbeta}
\end{figure}

Fig.~\ref{fig:outliers_derivs_modbeta} reveals a separation between the two groups in the $\mathrm{dlog}\rho$-$\mathrm{dlog}T$ plane, with the derivatives computed at $r_{\rm HSE, 500c}$. Moving along the thin diagonal lines, the bias would change due to a different temperature at a fixed sum of derivatives (i.e. a fixed pressure derivative, see Eq.~\ref{eq:hs_mass}), whereas moving perpendicular to those lines the bias changes at fixed temperature, but due to changing derivatives.

\begin{figure*}
    \centering
    \includegraphics[width = \linewidth]{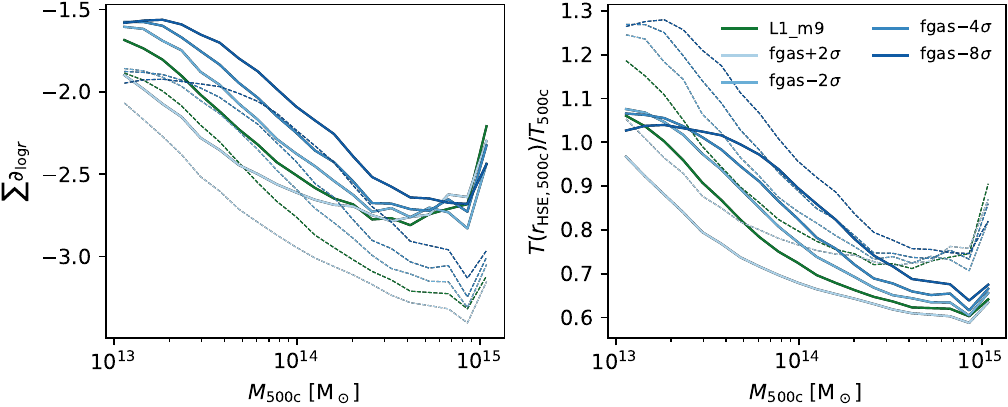}
    \caption{Medians of components of the hydrostatic mass equation (Eq.~\ref{eq:hs_mass}) for different FLAMINGO gas fraction variation models. The solid lines in the left panel show the sum of the logarithmic derivatives of the density and temperature profiles at $r_{\rm HSE, 500c}$. The solid lines in the right panel show the temperature at that radius normalised by the virial expectation ($T_{\rm 500c}$). The dashed lines indicate the required derivatives (or temperatures) given the temperature (or derivative) curve in the other panel. Different colours correspond to models with different cluster gas fractions, as indicated in the legend.}
    \label{fig:derivs_fits}
\end{figure*}

Objects with the largest hydrostatic mass (red in Fig.~\ref{fig:outliers_derivs_modbeta}) in each bin of true mass have more negative derivatives (i.e. steeper profiles) compared to those with the lowest hydrostatic mass (black in Fig.~\ref{fig:outliers_derivs_modbeta}). Qualitatively, this effect is independent of the true mass. Shallow profiles, characterized by small absolute values of the derivatives, may arise due to dynamical disturbance within clusters. We investigated this by examining correlations between bias and the center of potential - center of mass offset, a measure for relaxedness \citep{Evrard1993, Mohr1993}, but found no correlations (see Appendix \ref{sec:dynamical_state} for details). Previous work by \citet{Biffi2016} found a slight dependence of the bias on dynamical state, but we note that those authors did not match the goodness-of-fit and that their sample was very small, including just 6 regular and 8 disturbed systems.

The difference in the derivatives between the high and low inferred hydrostatic mass subsamples is found for all methods of computing profile derivatives, both for the parametric fits and the derivatives obtained from cubic-spline interpolation.

From Fig.~\ref{fig:outliers_derivs_modbeta} it can be seen that the two subsets can be connected by moving almost perpendicularly to the thin lines, indicating that the change in inferred hydrostatic mass could be explained by a change in the pressure gradient at fixed temperature. This agrees with results from \citet{Ansarifard2020}, who provide a partial empircal correction to the hydrostatic mass bias by using the gradient of the density profile. To confirm this, we examine the temperatures at $r_{\rm HSE, 500c}$ for these two groups of outliers. Fig.~\ref{fig:outliers_temp_modbeta} illustrates that objects exhibiting low and high inferred hydrostatic masses have comparable temperatures within identical narrow true mass bins.
Consequently, we infer that the scatter in the bias at a fixed mass and at fixed goodness-of-fit can be accounted for by differences in the derivatives of the temperature and density profiles, as depicted in Fig.~\ref{fig:outliers_derivs_modbeta}, indicating steeper (shallower) pressure profiles for the largest (smallest) hydrostatic masses at fixed true mass. 

\subsection{Components of the hydrostatic mass equation} \label{sec:origin_of_bias}

To further investigate the differences in the hydrostatic mass bias at fixed true mass between the different FLAMINGO model variations, we analyze the constituent components of the hydrostatic mass equation and their expected values in a virialised system. There are three variable parts in Eq.~\ref{eq:hs_mass}, the temperature $T(r_{\rm HSE, 500c})$, the sum of derivatives 
\begin{equation}
    \sum\partial_{\mathrm{log} r} (r_{\rm HSE, 500c}) = \left[ \frac{\mathrm{dlog}T}{\mathrm{dlog}r} + \frac{\mathrm{dlog}\rho}{\mathrm{dlog}r}\right] \, ,
\end{equation} 
and the radius $r_{\rm HSE, 500c}$. At fixed $M_{\rm 500c}$ the variation in the radius $r_{\rm HSE, 500c}$ across different physics model variations is minimal, so we do not delve into it any further. The temperature $T$ and the logarithmic derivative of the pressure $\sum\partial_{\mathrm{log} r}$  could exhibit significant variation between the different models, and hold distinct expectations for self-similar virialised haloes, to which we can compare.

At the spherical overdensity radius $r_{\rm 500c}$, the temperature of a self-similar virialised halo has the expected value \citep[for a derivation see e.g.][]{Braspenning2023}
\begin{equation} \label{eq:T500}
    T_\text{500c} = \frac{\mu m_{\rm p}}{2 k_{\rm B}} \left( \frac{500 G^2}{2} \right)^{1/3} M_{\rm 500c}^{2/3} H(z)^{2/3} \, .
\end{equation}
Additionally, in hydrostatic equilibrium, an isothermal sphere has a logarithmic density slope $\frac{d\log \rho}{d \log r} = -2$. While the equation for hydrostatic mass accommodates for a non-constant temperature profile, the sum of the derivatives should still equal $-2$ if the temperature equals $T_{\rm 500c}$ at $r_{\rm 500c}$\footnote{This follows from entering $T_{\rm 500c}$ for the temperature and demanding no bias ($M_{\rm HSE, 500c} = M_{\rm 500c}$).}. However, \citet{Braspenning2023} have shown that FLAMINGO haloes are not self-similar.

Fig.~\ref{fig:derivs_fits} shows the median sum of derivatives in the left panel, and in the right panel the median temperature divided by the virial expectation $T_{\rm 500c}$. The right panel shows that the deviation of the temperature from $T_{\rm 500c}$ is strongly mass dependent, and that there is a strong model dependence. Lower gas fractions models yield higher temperatures. At the lowest masses considered ($<2 \times 10^{13} ~\mathrm{M_\odot}$) the temperatures are similar.

The left panel shows differences in the sum of derivatives. The gas fraction variations show similar relative offsets as in the right panel. The dashed lines indicate, at each mass, the derivatives that would be required to have no bias (i.e. $b_{\rm HSE} = 0$) given the temperatures in the right panel. In the right panel, all solid lines are always below their corresponding dashed line, indicating that the total thermal force is insufficient to obtain a zero bias.

In the top left panel of Fig.~\ref{fig:variations} in Section \ref{sec:model_variations}, a crossing point is seen when we compare the bias for different gas fractions. At low mass, lower gas fractions give a stronger bias, whereas at high mass the inverse is true. This can be understood in terms of the different shapes of the derivatives and temperatures as a function of mass. Towards low masses ($<7 \times 10^{13} ~\mathrm{M_\odot}$) both the derivatives and temperature as a function of mass flatten. However, the temperatures flatten earlier than the derivatives, with the inflection happening at increasingly higher masses for lower gas fractions. The different inflection points for derivatives and temperatures result in a rapidly changing bias with mass, until the point where both curves flatten, which happens at $2 \times 10^{13} ~\mathrm{M_\odot}$ for fgas$-8\sigma$. The offset in bias values at higher masses ($>10^{14} ~\mathrm{M_\odot}$) can be understood because the distance between the solid and dashed lines is smaller for the lower gas fraction variations. This implies that they are closer to full thermal pressure support. 

The derivatives become insensitive to mass above $3 \times 10^{14} ~\mathrm{M_\odot}$, whereas the temperatures for the most extreme gas fraction variations become insensitive to mass towards the lowest masses ($<3 \times 10^{13} ~\mathrm{M_\odot}$).

\subsection{Energy balance} \label{sec:energy_balance}

\begin{figure}
    \centering
    \includegraphics{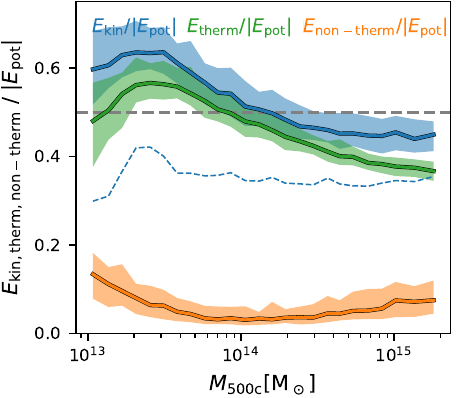}
    \caption{Ratio between the total kinetic (blue), thermal (green), and non-thermal (orange) energy to the potential energy within $r_{\rm 500c}$ as a function of the true mass. Solid lines show the medians, and the shaded regions show the $16^{\rm th}$ to $84^{\rm th}$ percentiles. Below $10^{14}~\mathrm{M_\odot}$ there is an excess of kinetic energy, whereas at higher masses there is a deficit compared to the value expected from the virial theorem, indicated by the grey dashed horizontal line. The contribution from non-thermal kinetic energy reaches 20 per cent both for low and high masses, but decreases to 5 per cent for intermediate masses. The dashed blue line indicates the total kinetic energy after applying a correction for the external thermal pressure. When external pressure is accounted for, clusters of all masses are sub-virial at $r_\text{500c}$.}
    \label{fig:energy_ratios}
\end{figure}

Non-thermal pressure has been posited as one solution to the discrepancy between hydrostatic mass and true mass. Because under the assumption of hydrostatic equilibrium, the mass is computed by equating the gravitational force with the thermal pressure gradient, if there is an additional pressure gradient from the non-thermal component, the gravitational force and hence the mass would be underestimated. Similarly, if there is significant net rotation, then the centrifugal force would lead to an underestimate of the mass. Furthermore, there is no guarantee that the assumption of hydrostatic equilibrium is valid.

Since deviations from hydrostatic equilibrium are present in our sample, we study whether these can be understood in terms of deviations from virial equilibrium. According to the equations of virial equilibrium, the gravitational potential energy of the gas should equal $-2$ times its kinetic energy. Following the derivation of virial equilibrium in the presence of external pressure \citep[e.g.][]{draine_ism}
\begin{align}
    E_{\rm pot} &= - 2E_{\rm kin} \\
    &= -2 \left( E_{\rm non-therm} + E_{\rm therm}  - E_{\rm therm, 0} \right) \, ,
\end{align}
where $E_{\rm therm, 0}$ represents a correction to the thermal energy due to external thermal pressure,
\begin{equation}
    E_{\rm pot} = -2E_{\rm non-therm} - 3 \int P \mathrm{d}V + \oint \mathrm{d}\Vec{S} \cdot \Vec{r} P \, ,
\end{equation}
where $\mathrm{d}\vec{S}$ is the surface element, and we neglect the non-thermal component of the external pressure. 
The first integral term is the thermal energy within the volume under consideration
\begin{equation}
    E_{\rm therm} = \frac{3}{2} \int P \mathrm{d}V \,
\end{equation}
where the factor $3/2$ follows from the equipartition theorem for a monatomic gas, and the second integral term expresses the pressure exerted on the surface within which the thermal energy is computed. Assuming the external pressure exerted on the spherical surface is a constant, $p_0$, we can write
\begin{equation}
    E_{\rm therm, 0} = \frac{1}{2} \oint \mathrm{d}\Vec{S} \cdot \Vec{r} P = \frac{3}{2} p_0 V \, .
\end{equation}
We compute $p_0$ as the volume weighted mean thermal pressure at $r_{\rm 500c}$. We find that at $r_\text{500c}$, the surface pressure yields a $\approx$ 20 to 40 per cent correction to the thermal energy required for equilibrium. This shows that the external pressure is an important effect when considering deviations from virial equilibrium at $r_{\rm 500c}$.

The potential energy is computed by summing the contributions of the potential energy between the gas particles and all particles (gas, stars, and dark matter) within $r_{\rm 500c}$
\begin{equation}
    E_{\rm pot} = \sum_{i \in \mathrm{gas}} \sum_{j \in \mathrm{all}} - \frac{Gm_i m_j}{r_{ij}} \, ,
\end{equation}
where $m_i$ and $m_j$ denote the masses of particles $i$ and $j$, respectively, $r_{ij}$ represents the distance between the two particles, and pairs are only counted once. The kinetic terms are defined as
\begin{align}
    E_{\rm non-therm} &= \sum_{i \in \mathrm{gas}} \frac{1}{2} m_i v_i^2 \, ,\\
    E_{\rm therm} &= \frac{3} {2} \sum_{i \in \mathrm{gas}} N_i\mathrm{k_B}T_i  \, ,
\end{align}
where $v_i$ represents the magnitude of the peculiar velocity of particle $i$ with respect to the center of mass velocity of all matter within $r_{\rm 500c}$ (we neglect the Hubble flow inside $r_\text{500c}$), $T_i$ the temperature, and $N_i = \frac{m_{{\rm gas},i}}{\mu m_{\rm H}}$. We compute this last value by dividing the mass of simulation gas particle $i$ by the mean particle mass, $\mu m_\text{H}$, where we assume $\mu = 0.6$, which is appropriate for a fully ionized plasma. Note that the non-thermal energy includes contributions from turbulence, bulk motions,  and rotation.

Fig.~\ref{fig:energy_ratios} shows the ratios of the non-thermal ($E_{\rm non-therm}$; orange), thermal ($E_{\rm therm}$; green), and total kinetic (without surface term) ($E_{\rm kin}$; blue) energies to the potential energy ($E_{\rm pot}$) as a function of the true mass ($M_\text{500c})$. The dashed grey horizontal line indicates the expected ratio between kinetic and potential energy of 0.5 from the virial theorem. Below $3 \times 10^{14}~\mathrm{M_\odot}$ there is an excess of kinetic energy of $\approx 20$ per cent, while at higher masses there is a lack of kinetic energy, gradually growing towards larger masses, reaching 10 per cent for the highest masses, $2 \times M_{\rm 500c} \sim 10^{15} ~ \mathrm{M_\odot}$. 

The dashed blue line indicates the value for the total kinetic energy if the external pressure correction is applied. For all masses, this correction yields a deficiency of kinetic energy relative to the virial expectation, which suggests strong pressure confinement by the surrounding medium, i.e. the gas just outside $r_\text{500c}$. Hence, even though at low mass, the total kinetic energy exceeds the virial expectation, the gas is sub-virial if the external pressure is taken into account. We note that \citet{Shaw2006} found a 15 per cent correction to virial equilibrium when using the surface pressure at $r_{\rm 178c}$ for dark matter haloes. Though our correction is at most twice as large, we include hydrodynamics and measure at a significantly smaller radius.

We note that the external pressure contribution was computed under the assumption of constant pressure, equal to the volume weighted average, on the spherical surface at $r_{\rm 500c}$. In reality, the pressure is unlikely to be azimuthally symmetric due to the presence of substructure and, at low mass, outflows driven by feedback. We neglected external ram pressure, which would be negative in the case of outflows. Although the magnitude of the correction for external pressure is thus uncertain, the deficiency in kinetic energy relative to the virial expectation when using the external pressure term suggests that significant work would have to be done to move the gas beyond $r_{\rm 500c}$.

We can understand the excess of kinetic energy below masses of $3\times 10^{14}~\mathrm{M_\odot}$ to be due to the injection of kinetic energy from baryonic feedback. At the highest masses the baryon fractions are close to universal \citep{Schaye2023} and cooling flows are more prevalent \citep{Braspenning2023}, which could reduce the thermal energy relative to virial expectations, in line with what we show in Fig.~ \ref{fig:energy_ratios}. The shortage of kinetic energy at the highest masses may also imply that these systems are still in the process of virialization, even at $r_{\rm 500c}$.

Non-thermal motions account for 5 to 15 per cent of the total kinetic energy, the contribution is smallest around $10^{14}~\mathrm{M_\odot}$, and increases towards both lower and higher masses. Since we ignored turbulence, rotation, and bulk motions in the hydrostatic mass equation (Eq.~\ref{eq:hs_mass}), we may expect a bias of about $b_{\rm HSE} \approx -0.1$ if the gradient of the non-thermal forces follows those of the thermal pressure, which is a bias value close to what we find at low mass for our fiducial X-ray luminosity weighted thermodynamic profiles (see Fig.~\ref{fig:fit_options_r500hs}). However, the mass bias measured from volume weighted profiles is smaller than the fraction of the non-thermal kinetic energy (see Fig.~\ref{fig:hs_bias_weight_comparison}), though their mass dependencies have similar shapes. This could imply that the turbulent energy fraction is larger towards the centre than at $R_\text{500c}$ where we measure the bias. 

The scatter of $\sim 10$ per cent in Fig.~\ref{fig:energy_ratios} is in line with the scatter in the hydrostatic bias we find for $M_{\rm 500c} > 3 \times 10^{13} ~ \mathrm{M_\odot}$ (see Fig.~\ref{fig:hs_bias_weight_comparison}), but we do not observe the steep increase in scatter at the lowest masses seen for the bias.

\citet{Martizzi2016}, who simulated clusters in the $M_{\rm 500c} \approx 1-3 \times 10^{14}~\mathrm{M_\odot}$ range, found that, using volume weighting, at $r_{\rm 500c}$ non-thermal energy can fully account for the hydrostatic bias. We find that in this mass range, the non-thermal energy in FLAMINGO is indeed sufficient to achieve virial equilibrium if we ignore external pressure. Our estimate of $\approx 10$ per cent non-thermal energy is furthermore in line with what is found in observations by \citet{Ettori2022} for massive clusters. 

Finally, we note that several sources of non-thermal energy channels are not explored in this work, such as magnetic fields and cosmic rays, which could potentially contribute significantly \citep[see e.g.][]{Rusdkowski2023}. We note that since these processes are not included in FLAMINGO, haloes are expected to reach energy equilibrium without including terms representing their contributions.

\subsection{Mass decomposition from the Euler equation}
An alternative approach to using hydrostatic equilibrium in estimating the mass is to find the mass contained within a sphere using the Euler equation, rewritten with the Poisson equation and Gauss' theorem,
\begin{align} \label{eq:euler}
    M_{\rm tot} &= \frac{1}{4\pi G} \int_{\partial V} \mathrm{d}\vec{S}~ \cdot \vec{\nabla}\phi , \\
    &=  \frac{1}{4\pi G} \int_{\partial V} \mathrm{d}\vec{S}~ \cdot \left[ -\frac{1}{\rho_{\rm gas}} \vec{\nabla} P_{\rm therm} - (\vec{v}\cdot \vec{\nabla})\vec{v} - \frac{\partial \vec{v}}{\partial t} \right] \, ,
\end{align}
with $\phi$ the gravitational potential. When every term is accounted for, this is guaranteed to yield the total mass, hence it can be used to understand which terms are important, aside from the thermal pressure gradient.
Eq. \ref{eq:euler} does not assume spherical symmetry or equilibrium, and will always return the total mass contained within the sphere. Following \citet{Suto2013} (see also \citealt{Oppenheimer2018}), we decompose the mass into four effective terms in spherical coordinates:
\begin{align}
    &M_{\rm therm} = - \frac{1}{4\pi G} \int_{\partial V} \mathrm{d} S~ \frac{1}{\rho_{\rm gas}} \frac{\partial P_{\rm therm}}{\partial r} , \label{eq:Mtherm} \\ 
    &M_{\rm rot} = \frac{1}{4\pi G} \int_{\partial V} \mathrm{d} S~ \frac{v_{\theta}^2 + v_{\phi}^2}{r} , \\
    &M_{\rm stream} = -\frac{1}{4\pi G} \int_{\partial V} \mathrm{d}S~ \left[ v_r \frac{\partial v_r}{\partial r} + \frac{v_{\theta}}{r}\frac{\partial v_r}{\partial\theta} + \frac{v_{\phi}}{r \sin \theta}\frac{\partial v_r}{\partial \phi} \right] ,\\
    &M_{\rm accel} = -\frac{1}{4\pi G} \int_{\partial V} \mathrm{d}S~ \frac{\partial v_r}{\partial t} .
\end{align}
We evaluate the equations by creating $10\times10$ angular bins in a narrow radial bin, $[0.95 - 1.05] \times r_{\rm 500c}$. For each bin, we compute volume weighted and X-ray luminosity weighted averaged densities, pressures and velocities, angular derivatives are found in a discrete fashion between neighbouring angular bins. The acceleration term cannot be directly computed from simulation output as those lack the required fine time resolution. However, because, by definition, the sum of the four mass terms equals the total mass, it can be identified as the remaining mass once the other three terms are accounted for
\begin{equation}
    M_{\rm rem} = M_{\rm accel} = M_{\rm tot} - \left(M_{\rm therm} + M_{\rm rot} + M_{\rm stream} \right). \label{eq:Mrest}
\end{equation}
Note that except for $M_\text{rot}$, the mass terms can be negative, which happens if they do not counteract gravity. 

The averaging of velocities in angular bins erases the small-scale turbulence present in the system. Hence, the remainder mass term will also include contributions from such turbulent motions.

The approach in this section makes fewer assumptions than using the virial theorem. In virial equilibrium, the acceleration vanishes, and the remainder term $M_{\rm rem}$ would account only for those turbulent motions that are averaged out when computing the velocities in angular bins. Furthermore, for the computation of the hydrostatic mass, we assume spherical symmetry, hence the angular derivatives $\frac{\partial v_r}{\partial \theta}$ and $\frac{\partial v_r}{\partial \phi}$ vanish. For pure pressure support, the radial derivative of the velocity vanishes $\frac{\partial v_r}{\partial r} = 0$, and there is no angular motion $v_\theta = v_\phi = 0$. Taking all those narrowing assumptions together, $M_{\rm rot} = M_{\rm stream} = M_{\rm accel} = 0$, and the only remaining term is $M_{\rm therm}$, which is what is usually considered when computing hydrostatic equilibrium masses.

\begin{figure*}
    \centering
    \includegraphics[width=.49\linewidth]{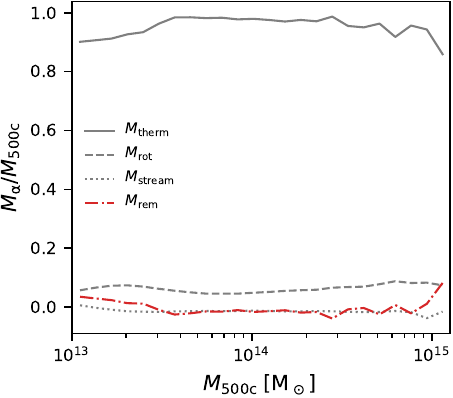}
    \includegraphics[width=.49\linewidth]{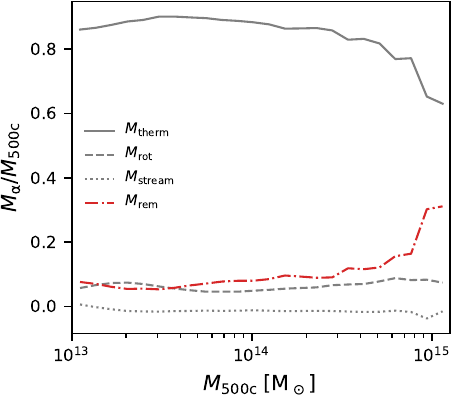}
    \caption{Contributions of different terms (different line styles; see equations \ref{eq:Mtherm} - \ref{eq:Mrest}) resulting from a decomposition of the Euler equation towards the total mass $M_{\rm 500c}$. The left and right panels show volume weighted and X-ray weighted results, respectively. The calculations were done at $r_\text{500c}$ in $10\times10$ angular bins, see the text for details. The thermal component, representing the thermal pressure gradient, dominates. This is followed, in the case of volume weighting, by the term representing rotational support. The contribution of streaming motions is negligible at all true total masses. The remainder mass term is equal to the difference between the total mass and the other mass terms, it represents turbulent motions that were averaged out due to the angular binning as well as acceleration, its contribution is small at all masses for volume weighting, but increases for the highest masses if we use X-ray weighting.}
    \label{fig:mass_terms}
\end{figure*}

Fig.~\ref{fig:mass_terms} shows the fractional contributions of the various terms as a function of the true total mass for the case of volume weighting (left panel) and X-ray luminosity weighting (right panel). Focusing first on the left panel, we see that the thermal mass, equivalent to the hydrostatic equilibrium mass, dominates across the entire mass range. The streaming mass is negligible. However, the mass due to rotational motion contributes $\sim 10$ per cent of the total mass. Finally, the remaining few per cent is made up of the $M_{\rm rem}$ term, which includes turbulence (due to the angular binning applied in the calculation of the rotational and streaming masses) and the acceleration term. 

In Fig.~\ref{fig:energy_ratios} we have shown that the non-thermal energy fraction varies smoothly with mass and has a slight uptick for the lowest masses. That finding is consistent with the slight increase in remainder mass below $4\times 10^{13}~\mathrm{M_\odot}$ we find here.

The right panel of Fig.~\ref{fig:mass_terms} shows that if we use X-ray weighting to compute averages in the angular bins used for the Euler equations, we find a sharp upturn in remainder mass at the very highest true masses ($M_{\rm 500c} > 10^{15} ~\mathrm{M_\odot}$). This upturn can be due to increased turbulence in the X-ray gas that is averaged out in the angular binning or it can be due to the acceleration term. Such an increased acceleration could indicate a lack of virialisation of the gas dominating the X-ray flux or a large-scale cooling-flow for those highest-mass objects, and may serve as an explanation for the increased hydrostatic bias above $M_{\rm 500c} = 10^{15}~\mathrm{M_\odot}$ found in this work from X-ray weighted profiles (see Fig.~\ref{fig:hs_bias_weight_comparison}). That aligns with the significant velocity gradients near $r_{\rm 500c}$ for the most massive clusters in FLAMINGO \citep{Kay2024} for X-ray weighted gas.

The most massive haloes tend to be the least relaxed and might have significant dynamical disturbances. As a result, they can be less axisymmetric and perhaps more clumpy, which are the conditions under which the largest differences between weighting schemes are expected.

\section{Conclusions} \label{sec:conclusion}

This study examined the hydrostatic mass bias for galaxy clusters from the FLAMINGO cosmological hydrodynamical simulation suite. Assuming hydrostatic equilibrium, hydrostatic masses are computed using the derivatives of the X-ray luminosity-weighted density and temperature profiles at $r_{\rm HSE, 500c}$, along with the absolute temperature at that radius (Eq.~\ref{eq:hs_mass}). Our main results can be summarized as follows:
\begin{itemize}
    \item As a function of the true mass, the median hydrostatic bias $b_\text{HSE}= (M_\text{HSE,500c}-M_\text{500c})/M_\text{500c} \approx -0.1$ at group mass scales, but increases to $\approx -0.2$ at the cluster mass scale (Fig.~\ref{fig:fit_options_r500hs}, left panel).  The scatter in the bias is smallest for $M_\text{500c}\sim 10^{14}\,\text{M}_\odot$, where $\sigma(b_{\rm HSE})\approx 0.1$. The scatter rapidly diverges for $M_\text{500c} \ll 10^{14}\,\text{M}_\odot$ and also increases markedly for $M_\text{500c} \gg 10^{14}\,\text{M}_\odot$. The results for the median bias are in good agreement with the bias inferred from comparisons between hydrostatic masses from X-ray observations and weak-lensing masses, though the observations are not yet very constraining due to the large differences between different studies. The predictions also align with results from other simulations.

    \item Although theoretical studies conventionally predict the hydrostatic mass bias as a function of the true mass, to correct observed hydrostatic masses, we require the bias as a function of the inferred hydrostatic mass. We predict the median bias to be nearly independent of the hydrostatic mass (Fig.~\ref{fig:fit_options_r500hs}, right panel). The qualitative difference with the bias as a function of true mass is caused by the upscatter of objects with a lower true mass. At fixed hydrostatic mass these lower-mass objects mix with slightly more massive objects whose mass is more strongly underestimated, thus erasing the mass dependence of the bias.

    \item The hydrostatic mass bias is highly sensitive to the weighting applied to particles when computing the spherically averaged thermodynamic profiles (Fig.~\ref{fig:hs_bias_weight_comparison}). Compared with X-ray luminosity weighted profiles, volume weighted profiles, which are theoretically preferable but cannot be obtained observationally, result in a much small hydrostatic mass bias for $M_\text{500c} \gg 10^{13}~\text{M}_\odot$. This suggests that much of the observed bias is not due to deviations from thermal hydrostatic equilibrium, but is caused by the biased nature of X-ray emitting gas. 

    \item The hydrostatic mass bias is insensitive to the assumed functional form of the density profile (Fig.~\ref{fig:fit_options_r500hs}).

    \item The FLAMINGO suite includes simulations for which the subgrid stellar and AGN feedback prescriptions have been re-calibrated to fit perturbed versions of the observed cluster gas fractions or galaxy stellar mass function. The models varying the cluster gas fractions and/or galaxy stellar mass function exhibit up to a factor 2 offset in the hydrostatic bias. Compared with our fiducial simulation, models with lower gas fractions yield less biased hydrostatic masses for $M_\mathrm{500c} > 10^{14}~\mathrm{M_\odot}$, but predict a stronger bias at lower masses (Figs.~\ref{fig:variations} and \ref{fig:variations_hse}).

    \item At higher redshifts the median bias is a more steeply declining function of the true mass. For $M_\text{500c}\sim 10^{15}\,\text{M}_\odot$ the true mass is more strongly underestimated at higher redshift, but for $M_\text{500c}\sim 10^{13}\,\text{M}_\odot$ $b_\text{HSE}$ increases with redshift, leading to an overestimate of the true mass for $z>1$ (Fig.~\ref{fig:redshift}). The median bias is insensitive to the hydrostatic mass at all redshifts, but $b_\text{HSE}$ increases for $z\gtrsim 1$. The scatter in the bias increases with redshift, regardless of whether we bin in true or hydrostatic mass.  

    \item The magnitude of the bias depends on how well the adopted analytic functions describe the actual density and temperature profiles of the cluster (Fig.~\ref{fig:gof_biases}). For example, at a true or hydrostatic mass of $10^{14}\,\text{M}_\odot$ the median bias strengthens from $b_{\rm HSE} = -0.1$ to $b_{\rm HSE} = -0.3$ from the best 50 per cent to the worst 10 per cent fits to the density and temperature profiles. Poorer fits also yield a more strongly mass-dependent bias. The scatter in the bias more than doubles from the best to the worst fits. 

    \item The scatter in hydrostatic bias at fixed true mass can be explained by differences in the pressure gradients between haloes, but not by differences in the temperatures. Haloes with a higher value for the hydrostatic bias have similar temperatures, but steeper pressure gradients, compared to objects with a lower value of $b_\text{HSE}$ (Figs.~\ref{fig:outliers_derivs_modbeta} and \ref{fig:outliers_temp_modbeta}).

    \item At fixed true mass, the temperature and pressure gradients differ strongly between the FLAMINGO models with different cluster gas fractions (Fig.~\ref{fig:derivs_fits} ). The clusters in models with smaller gas fractions have shallower pressure gradients, but higher temperatures.

    \item For true masses of $M_{\rm 500c} \lesssim 10^{14}~\mathrm{M_\odot}$ the sum of the thermal and non-thermal energies exceeds what is required for virial equilibrium (Fig.~\ref{fig:energy_ratios}). At higher masses,  the total kinetic energy falls increasingly short of what is required for virial equilibrium, suggesting that the most massive objects are still in the process of virialization even at $r_{\rm 500c}$. However, if we account for the external pressure at $r_\text{500c}$, then the total kinetic energy is sub-virial at all masses.
    
    \item Non-thermal energy contributes $\sim 10-20$ per cent to the total kinetic energy (Fig.~\ref{fig:energy_ratios}). Naively, this could be taken to imply that the hydrostatic mass bias could be due to the neglect of the non-thermal energy. However, the volume weighted mass bias is much smaller than the X-ray weighted mass bias. This suggests that the non-thermal energy is concentrated towards the cluster centre and hence has relatively little effect on the hydrostatic mass measured at $R_\text{500c}$.

    \item We quantify the contribution of different terms in the Euler equation to the total mass (Fig.~\ref{fig:mass_terms}). Coherent rotational support is $\sim 10$ per cent at all true masses. When using volume weighting, the acceleration term, which includes small-scale turbulence averaged out by binning, contributes $\sim 10$ per cent at the lowest masses but quickly becomes negligible with increasing true mass. If, instead, X-ray weighting is used, the acceleration term contributes up to 30 per cent of the total mass for the most massive objects. This hints at incomplete virialisation of the gas dominating the X-ray emission for those masses and provides a possible explanation for the increased hydrostatic bias for the highest mass objects.

\end{itemize}

By comparing different FLAMINGO simulations, we have shown that the hydrostatic mass bias is strongly dependent on the median cluster gas fractions and thus on the strength of stellar and AGN feedback. Because the differences between reasonable gas fraction models is larger than the error bars on many cluster mass measurements, tests of observational methods should thus use hydrodynamical simulations that span a range of gas fractions. This is also critical for estimating errors on X-ray inferred masses for cluster cosmology. On top of that, the strong systematic dependence of the bias on the quality of the fit to the density and temperature profiles potentially creates another source of significant systematic uncertainty that should be accounted for.

As a result of the magnitude and mass dependence of the scatter, the dependence of the hydrostatic bias on mass is rather different depending on whether the clusters are binned in terms of their true or hydrostatic masses. This means that different values for the bias are required when attempting to correct hydrostatic masses inferred from X-ray observations and when attempting to impose a hydrostatic bias on the true masses predicted by simulations. 

Taken together, our results suggest the need to employ forward modeling using hydrodynamical simulations when precise cluster masses are required. The simulations should have volumes sufficiently large to model the full distribution of the bias at fixed mass because this is required to model the differential effects of upscatter and downscatter between mass bins as well as to model selection effects. The simulations should vary the uncertain subgrid prescriptions for feedback processes within the ranges allowed by observation. 

While we used X-ray luminosity-weighted thermodynamic profiles, in this work, we did not attempt to create virtual observations or to account for the limitations of specific surveys. Our conclusions could be strengthened by the production of mock X-ray data and by mimicking observational techniques. This is particularly important in the light of the fact that we infer a much smaller hydrostatic mass bias from volume weighted average thermodynamic profiles than when we use X-ray luminosity weighting. 

\section*{Acknowledgements}
We acknowledge support from research programme Athena 184.034.002 from the Dutch Research Council (NWO).
This work used the DiRAC@Durham facility managed by the Institute for Computational Cosmology on behalf of the STFC DiRAC HPC Facility (www.dirac.ac.uk). The equipment was funded by BEIS capital funding via STFC capital grants ST/K00042X/1, ST/P002293/1, ST/R002371/1 and ST/S002502/1, Durham University and STFC operations grant ST/R000832/1. DiRAC is part of the National e-Infrastructure. This project has received funding from the European Research Council (ERC) under the European Union’s Horizon 2020 research and innovation programme (grant agreement No 769130).

\section*{Data Availability}
The data underlying the plots within this article are available on
reasonable request to the corresponding author. The FLAMINGO simulation data will eventually be made publicly available, though we note that the data volume (several petabytes) may prohibit us from simply placing the raw data on a server. In the meantime, people interested in using the simulations are encouraged to contact the corresponding author



\bibliographystyle{mnras}
\bibliography{joey} 

\appendix

\section{The effect of subhaloes on the hydrostatic bias} \label{sec:subhaloes}
\begin{figure}
    \centering
    \includegraphics{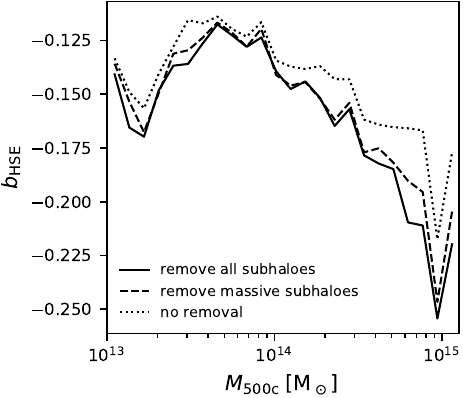}
    \caption{Median hydrostatic mass bias as a function of true mass obtained using the cubic-spline measurement approach for different exclusion criteria for subhaloes. The difference between removing all subhaloes or only the most massive ($M_{\rm sub} > 0.1 M_{\rm 500c}$) subhaloes is small.}
    \label{fig:exclude_subs}
\end{figure}

In X-ray observations, bright substructures are typically removed before extracting thermodynamic profiles. Potentially, the subtraction of such substructure could influence the determination of the hydrostatic bias. Figure \ref{fig:exclude_subs} shows that this is at most a small effect. Removing all subhaloes, which is our fiducial approach, gives only small differences in the median bias compared with no removal, and below $M_{\rm 500c} \sim 10^{14} ~ \mathrm{M_\odot}$ the removal of subhaloes becomes irrelevant. For the most massive haloes $M_{\rm 500c} \sim 10^{15}~\mathrm{M_\odot}$ the offset in the bias is largest, $\Delta b_{\rm HSE} \approx 0.05$. However, the effect of removing only the most massive ($M_{\rm sub} > 0.1 M_{\rm 500c}$) subhaloes is consistent with that of removing all subhaloes. This shows that only the brightest subhaloes can significantly affect the thermodynamic profiles. Hence, the removal of bright substructures, as is done in observations, is sufficient for computing the hydrostatic mass. 

\section{The effect of the cluster dynamical state on the hydrostatic bias} \label{sec:dynamical_state}
Since the hydrostatic mass calculation relies on the assumption of hydrostatic equilibrium, a natural consequence is that dynamically disturbed systems, such as recent mergers, may not be adequately captured by this approach, possibly resulting in incorrect mass determinations. We use the centre-of-mass -- centre-of-potential offset normalised to $r_{\rm 500c}$ as a measure of relaxedness, which has been shown to accurately distinguish between dynamical disturbed and relaxed systems \citep{Evrard1993, Mohr1993}.

\begin{figure}
    \centering
    \includegraphics{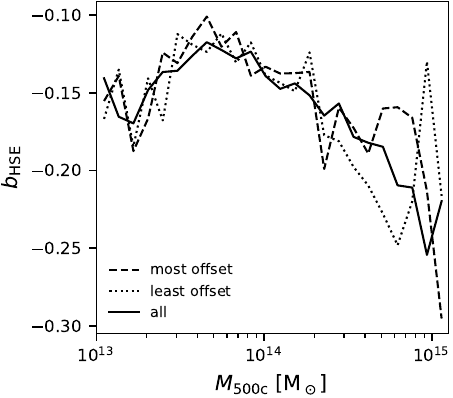}
    \caption{Median hydrostatic mass bias as a function of true mass obtained using the cubic-spline measurement approach for clusters in different dynamical states. The dashed line shows the median bias for the 10 per cent of haloes with the largest normalised center-of-mass -- center-of-potential offset (least relaxed), whereas the dotted line shows the median for the 10 per cent least offset (most relaxed) haloes. The difference between the most- and least offset haloes is very small, which implies that there is no significant correlation between bias and dynamical state.}
    \label{fig:relaxed_offsets}
\end{figure}

Figure \ref{fig:relaxed_offsets} shows that only for the most massive clusters there is a difference in bias between the 10 per cent most and 10 per cent least offset systems, and all are well within the scatter in the bias at a given mass. Clearly, relaxedness is not the driving cause of scatter in the bias and can be ignored for our purposes. If anything, the hydrostatic masses for the more relaxed (less offset) systems are more strongly biased compared to the less relaxed (more offset) systems. This result also holds when binning in hydrostatic mass instead of true mass. The effect of the dynamical state is negligible in both cases.

\section{Recalibrating FLAMINGO} \label{sec:recalibration}
During the calibration of the subgrid models for feedback in the FLAMINGO suite of simulations, a constant hydrostatic bias of $b_{\mathrm{HSE}} = -0.257$ was assumed based on comparisons between X-ray and weak lensing observations \citep{Kugel2023}. This affects the calibration of cluster gas fractions, by shifting X-ray inferred masses taken from the literature. In this work, we found that using an analytic fit to the cluster density profile, as would be done in the X-ray observations underlying the value of hydrostatic bias used in calibration, yields a mass-dependent bias (see Fig.~\ref{fig:fit_options_r500hs}). As a function of the inferred hydrostatic mass ($M_{\rm HSE, 500c}$) the bias is always smaller than the calibration value, and has only a slight mass dependence. 

\begin{figure}
    \centering
    \includegraphics[width = \linewidth]{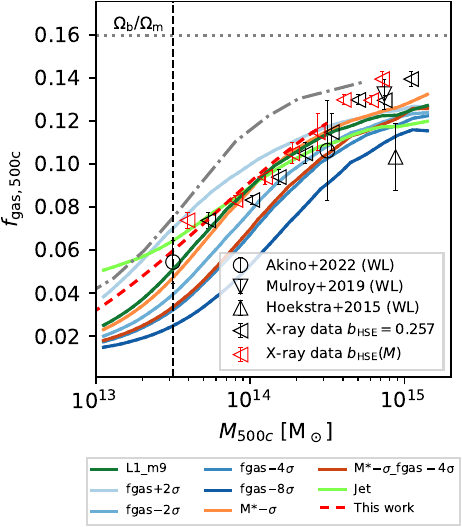}
    \caption{Median group and cluster gas fractions compared with observations. The red symbols indicate the X-ray observations if they are shifted with the $M_\text{HSE,500c}$-dependent hydrostatic bias found in this paper, whereas the black symbols use the constant hydrostatic bias of $b_\text{HSE}=-0.257$ assumed during the calibration of the FLAMINGO simulations.}
    \label{fig:gas_fractions}
\end{figure}

Fig.~\ref{fig:gas_fractions} shows the result of applying the $M_\text{HSE,500c}$-dependent hydrostatic bias from the modified-$\beta$ fit to the fiducial FLAMINGO model (see Fig.~\ref{fig:fit_options_r500hs}) to the observed gas fraction data. Red triangles show the new values, and the dashed red line is the prediction from the emulator of \citet{Kugel2023}. Calibrating the FLAMINGO model to this new data yields slightly different values for the 4 subgrid feedback parameters that are varied by the Gaussian process emulator. Table~\ref{tab:parameters} indicates that the AGN heating temperature, $\Delta T_\text{AGN}$, is in between those for the fiducial and the $+2\sigma$ gas fraction models. The black hole accretion rate boost factor, $\beta_\text{BH}$, is slightly larger than fiducial, but still significantly below the fgas$+2\sigma$ model. The stellar feedback couples slightly less efficiently to the ISM ($\fsn$ is lower) moving towards fgas$+2\sigma$, but the wind velocity is higher than in that model. 

Shifting the X-ray data with the $M_\text{HSE,500c}$-dependent hydrostatic bias, yields a unique set of parameters for which the emulator prediction for $b_\text{HSE}(M_\text{500c})$ exhibits a somewhat different shape and slope compared to the other gas fraction variation models, the enhancement in gas fraction at lower masses makes it more similar to the Jet model. The characteristic S-shape of the other models is absent and replaced by an almost linear increase in gas fraction with a weak flattening above $10^{14}~\mathrm{M_\odot}$. Even though both the AGN and supernova feedback are weaker in this model, and roughly halfway between the fiducial and fgas$+2\sigma$ model parameters, the gas fractions are close to the fiducial model across most of the mass range.

The general trend is that feedback becomes slightly weaker in this model, especially for low-mass groups where supernova feedback dominates over AGN feedback.
\begin{table}
    \centering
    \caption{Values of the calibrated subgrid parameters for the fiducial model and the $\pm 2\sigma$ feedback variations at intermediate resolution, compared to using the $M_\text{HSE,500c}$-dependent hydrostatic bias from this work (bottom row).}
    \label{tab:parameters}
    \begin{tabular}{llccl}
        \hline
       Prefix & $f_\text{SN}$ & $\vsn$ & $\tagn$ & $\betabh$ \\
    & & $(\kms)$ & (K)  \\
       \hline
       Fid\_m9             & 0.238 & 562 & $10^{7.95}$ & 0.514 \\
       fgas$+2\sigma$\_m9 & 0.219  & 577 & $10^{7.71}$ & 0.554 \\
       fgas$-2\sigma$\_m9 & 0.206  & 552 & $10^{8.08}$ & 0.497 \\
       This work & 0.228 & 584 & $10^{7.85}$ & 0.522 \\ 
    \hline
    \end{tabular}
\end{table}

\bsp	
\label{lastpage}
\end{document}